\documentclass[submission,copyright,creativecommons]{eptcs}
\providecommand{\event}{LSFA 2022}
\usepackage{underscore}
\usepackage[sort,nocompress]{cite}

\usepackage{microtype}
\usepackage{amssymb,amsfonts,amscd,amsbsy,mathtools}
\usepackage{bussproofs}
\usepackage[thmmarks,amsmath,standard]{ntheorem}
\usepackage{xparse}

\newenvironment{Thm}[1]{\begin{Theorem}#1}{\NoEndMark\end{Theorem}}
\newenvironment{Lem}[1]{\begin{Lemma}#1}{\NoEndMark\end{Lemma}}
\newenvironment{Def}[1]{\begin{Definition}#1}{\NoEndMark\end{Definition}}
\newenvironment{Exm}{\begin{Example}}{\NoEndMark\end{Example}}
\newenvironment{Rem}{\begin{Remark}}{\NoEndMark\end{Remark}}

\usepackage{quiver}

\DeclareMathAlphabet{\mathcal}{OMS}{cmsy}{m}{n}

\usepackage{xcolor-material}
\usepackage{stmaryrd}

\usepackage{bm}

\usepackage[inline]{enumitem}

\newcommand{\sigFont}[1]{\mathsf{#1}}
\newcommand{\sortFont}[1]{\mathsf{#1}}

\newcommand{\csPairFont}[1]{\mathcal{#1}}
\newcommand{\compFont}[1]{\mathsf{#1}}

\newcommand{\consFont}[1]{\mathsf{#1}}
\newcommand{\defFont}[1]{\mathtt{#1}}

\newcommand{\metaFont}[1]{\mathtt{#1}}

\newcommand{\nat}{\sortFont{nat}}
\newcommand{\lst}{\sortFont{list}}

\newcommand{\cost}{\compFont{c}}
\newcommand{\size}{\compFont{s}}
\newcommand{\leng}{\compFont{l}}
\newcommand{\mmax}{\compFont{m}}

\newcommand{\zero}{\consFont{0}}
\newcommand{\nil}{\consFont{nil}}
\newcommand{\suc}{\consFont{s}}
\newcommand{\cons}{\consFont{cons}}

\newcommand{\minus}{\consFont{minus}}
\newcommand{\quot}{\consFont{quot}}

\newcommand{\add}{\defFont{add}}
\newcommand{\double}{\defFont{dbl}}

\newcommand{\append}{\defFont{append}}
\newcommand{\rev}{\defFont{rev}}
\newcommand{\sumList}{\defFont{sum}}

\newcommand{\asort}{\iota}
\newcommand{\bsort}{\kappa}

\newcommand{\atype}{\sigma}
\newcommand{\btype}{\tau}

\newcommand{\afun}{\sigFont{f}}
\newcommand{\bfun}{\sigFont{g}}

\newcommand{\avar}{x}
\newcommand{\bvar}{y}
\newcommand{\cvar}{z}
\newcommand{\aListVar}{q}
\newcommand{\bListVar}{l}

\newcommand{\aterm}{s}
\newcommand{\bterm}{t}
\newcommand{\cterm}{u}

\newcommand{\num}[1]{\mathsf{#1}}

\newcommand{\aDom}{\csPairFont{A}}

\newcommand{\costInt}[1]{\mathcal{C}_{#1}}
\newcommand{\costFInt}[1]{\mathcal{F}^\cost_{#1}}
\newcommand{\sizeInt}[1]{\mathcal{S}_{#1}}

\newcommand{\signature}{\mathcal{F}}

\newcommand{\dfdS}{\mathcal{D}}
\newcommand{\ctrS}{\mathcal{C}}
\newcommand{\var}{\mathcal{X}}
\newcommand{\terms}{T(\signature,\var)} 
\newcommand{\bTerms}{T_b(\signature)} 
\newcommand{\vars}[1]{\mathtt{vars}(#1)} 

\newcommand{\sortset}{\mathcal{B}}
\newcommand{\simpletypeset}{\mathcal{T}_\sortset}

\newcommand{\rules}{\mathcal{R}}
\newcommand{\Nat}{\mathbb{N}}
\newcommand{\NatSizeSet}[1]{\Nat^{\typecount{#1}}}

\newcommand{\rulesArrow}{\to}
\newcommand{\arrzR}{\rulesArrow_\rules}
\newcommand{\arrz}{\rulesArrow}

\newcommand{\interpret}[1]{\llbracket#1\rrbracket}
\newcommand{\typeinterpret}[1]{\llparenthesis#1\rrparenthesis}

\newcommand{\typecount}[1]{K[#1]}

\newcommand{\ainterpret}[1]{\llbracket#1\rrbracket_{\alpha}^{\funcinterpret{}}}
\newcommand{\funcinterpret}[1]{\mathcal{J}_{#1}}

\newcommand{\costGt}{>}
\newcommand{\costGe}{\geq}
\newcommand{\sizeGe}{\sqsupseteq}
\newcommand{\cartGt}{\succ}
\newcommand{\cartGe}{\succcurlyeq}

\newcommand{\pair}[1]{\left\langle#1\right\rangle}

\newcommand{\tuple}[1]{\bm{#1}}
\NewDocumentCommand{\typeVec}{m o}{
    \IfValueTF{#2}{
        {}^{#2}\bm{#1}
    }{
        \bm{#1}
    }
}

\newcommand{\fatlambda}{\bm{\lambda}}
\newcommand{\app}{\ }
\newcommand{\semApp}{\cdot}

\newcommand{\dom}{\metaFont{dom}}

\newcommand{\hasType}{\mathbin{:}}

\newcommand{\subtermR}{\unrhd}

\newcommand{\asub}{\gamma}

\newcommand{\unit}{\metaFont{unit}}
\newcommand{\unitE}{\mathfrak{U}}

\newcommand{\dht}[1]{\metaFont{dh}(#1)} 
\newcommand{\compFunc}{\metaFont{comp}}
\newcommand{\dc}{\metaFont{dc}_\rules}
\newcommand{\idc}{\metaFont{idc}_\rules}
\newcommand{\rc}{\metaFont{rc}_\rules}
\NewDocumentCommand{\irc}{o}{
    \IfValueTF{#1}{
        \metaFont{irc}_{\rules_{#1}}
    }{
        \metaFont{irc}_{\rules}
    }
}

\newcommand{\semAppFunc}{\bm{\metaFont{App}}}
\newcommand{\costWrapper}[1]{(#1)}

\newcommand{\ar}{\metaFont{ar}}

\newcommand{\arrtype}{\Rightarrow}
\newcommand{\arrfunc}{\longrightarrow}
\newcommand{\arrfuncwm}{\Longrightarrow}
\newcommand{\intKey}{\funcinterpret{\sortset}}

\newcommand{\floor}[1]{\left\lfloor#1\right\rfloor}

\newcommand{\bigO}[1]{\mathcal{O}\left(#1\right)}

\newcommand{\cs}{cost--size} 
\newcommand{\csbegin}{Cost--size} 
\newcommand{\cstitle}{Cost--Size} 

\newcommand{\orcid}[1]{
    \href{#1}{\includegraphics[height=9pt]{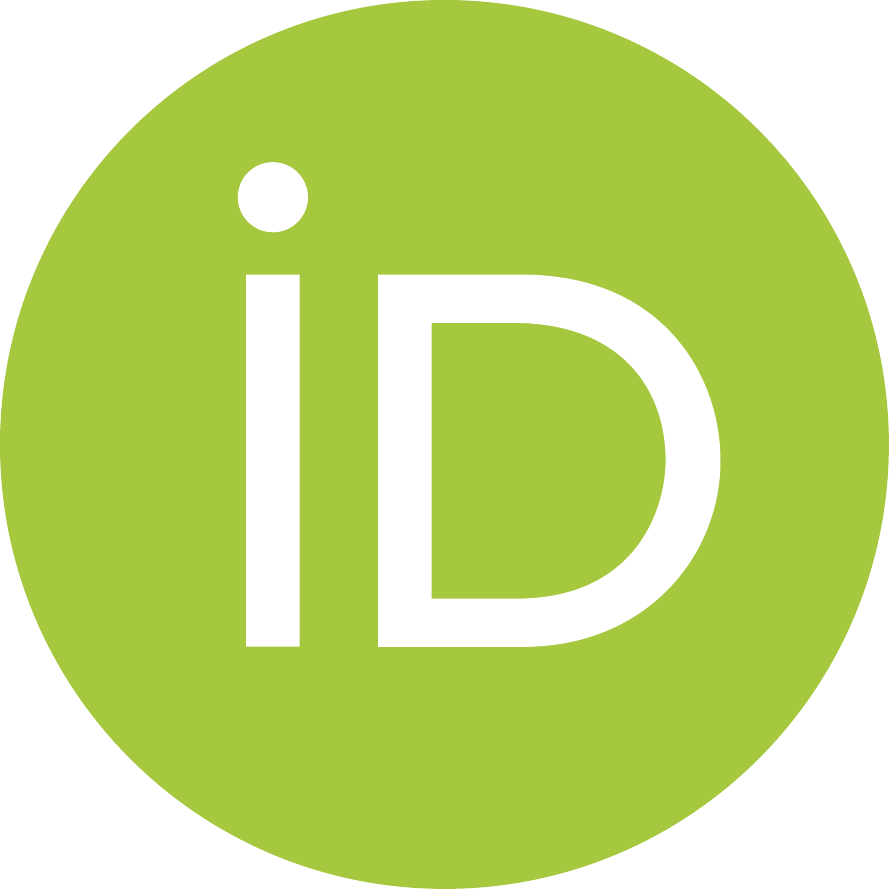}}
}

\title{Analyzing Innermost Runtime Complexity Through~Tuple~Interpretations}
\author{
    Liye Guo\thanks{
        The authors are supported by the
        NWO VIDI project ``CHORPE'',
        NWO VI.Vidi.193.075
        and the
        NWO TOP project ``ICHOR'',
        NWO 612.001.803/7571.
    }
    \orcid{https://orcid.org/0000-0002-3064-2691}
    \institute{Institute for Computing and Information Sciences\\
    Radboud University, The Netherlands
    }
    \email{l.guo@cs.ru.nl}
\and
    Deivid Vale\footnotemark[1]
    \orcid{https://orcid.org/0000-0003-1350-3478}
    \institute{Institute for Computing and Information Sciences\\
        Radboud University, The Netherlands
    }
    \email{deividvale@cs.ru.nl}
}
\def\titlerunning{
    Analyzing Innermost Runtime Complexity
    Through~Tuple~Interpretations
}
\def\authorrunning{L. Guo \& D. Vale}

\begin{document}
\maketitle

\begin{abstract}
    Time complexity in rewriting is naturally understood
    as the number of steps needed to reduce terms to normal forms.
    Establishing complexity bounds to this measure
    is a well-known problem in the rewriting community.
    A vast majority of techniques to find such bounds
    consist of modifying termination proofs
    in order to recover complexity information.
    This has been done for instance
    with semantic interpretations,
    recursive path orders,
    and dependency pairs.
    In this paper,
    we follow the same program by tailoring
    tuple interpretations to deal with innermost
    complexity analysis.
    A tuple interpretation interprets terms as tuples
    holding upper bounds to the cost of reduction
    and size of normal forms.
    In contrast with the full rewriting setting,
    the strongly monotonic requirement for cost components
    is dropped when reductions are innermost.
    This weakened requirement on cost tuples allows
    us to prove the innermost version of the
    compatibility result:
    if all rules in a term rewriting system
    can be strictly oriented,
    then the innermost rewrite relation is well-founded.
    We establish the necessary conditions for which
    tuple interpretations
    guarantee polynomial bounds
    to the runtime of compatible systems
    and describe a search procedure for such interpretations.
\end{abstract}

\section{Introduction}\label{sec:intro}

In the step-by-step computational model induced by rewriting,
time complexity is naturally understood as
the number of rewriting steps needed to reach normal forms.
Usually,
the cost of firing a redex
(i.e., performing a computational step)
is assumed constant.
So the intricacies of a low-level rewriting realization
(e.g., a concrete rewriting engine implementation) are ignored.
This assumption does not pose a problem as long as
the low-level time complexity needed to apply a rule is kept low.
Additionally, this abstract approach has the advantage
of being independent of the specific hardware platform
evaluating the rewriting system at hand.

In this rewriting setting,
a complexity function bounds the length
of rewrite sequences and is parametrized by the
size of the starting term of the derivation.
Two distinct complexity notions are commonly considered
in the literature:
derivational and runtime complexity,
and they differ by the restrictions imposed on the initial term of derivations.
On the one hand,
derivational complexity imposes no restriction
on the set of initial terms.
Intuitively,
it captures the worst-case behavior
of reducing a term to normal form.
On the other hand, runtime complexity requires
basic initial terms which,
conceptually,
are terms where a single function call is performed
on data (e.g., integers, lists, and trees) as arguments.

If programs are expressed by rewriting,
their execution time is closely related
to the runtime complexity of the associated rewrite system.
Similarly related are programs using call-by-value evaluation
strategy and innermost rewrite systems.
Therefore,
by combining these two concepts,
we obtain a connection between the cost analysis of
call-by-value programs and
the runtime complexity analysis of innermost term rewriting.
More importantly,
due to the abstract nature of rewriting,
it is feasible to forgo any specific programming language
detail and still derive useful term rewriting results
that may carry over to programs.
For an overview of the applicability of rewriting
to program complexity
the reader is referred to~\cite{moser:17,ava:mos:08}.

Therefore,
a rewriting approach to program complexity
allows us to fully concentrate on finding techniques
to establish bounds to the
derivational or runtime complexity functions.
A natural way to determine these bounds
is adapting the proof techniques
used to show termination
to deduce the complexity naturally induced by the method.
There is a myriad of works following this program.
To mention a few,
see~\cite{
    hof:lau:89,
    cichon:lescanne:92,
    baillot:lago:16,
    hof:01,
    bonfante:cichon:marion:touzet:01,
    moser:schnabl:wadmann:08
}
for interpretation methods,~\cite{hof:92,bonfante:et-al:01,weiermann:95}
for lexicographic and path orders,
and~\cite{nao:moser:08,noschinski:emmes:jurgen:11} for dependency pairs.
In this paper,
we follow the same idea and concentrate on
investigating the existence of upper bounds to the
innermost runtime complexity for applicative systems.
The termination method on which we base
our complexity analysis framework
upon is tuple~interpretations~\cite{kop:vale:21}.

Tuple interpretations are an instance of
the interpretation method.
Thus, we seek to interpret terms in such a way that
the rewrite relation can be embedded
in a well-founded ordering.
More precisely,
we choose an interpretation domain \( A \)
which is a set together with a well-founded order \( \costGt \) over \( A \)
and interpret terms as elements of \( A \) compositionally.
This interpretation of terms is such that
whenever a rewriting is fired, i.e.,
\( \aterm \arrz \bterm \),
the interpretations \( \interpret{\aterm} \) and \( \interpret{\bterm} \)
of \( \aterm \) and \( \bterm \) satisfy
\( \interpret{\aterm} \costGt \interpret{\bterm} \).
Hence,
a rewriting step on terms implies a strict decrease on \( A \).
The well-foundedness of such domains together with this
compatibility requirement on reduction
guarantee that no infinite reduction sequence on terms exists.

The defining characteristic of tuple interpretations is to allow for a split of
the complexity measure into abstract notions of cost and size.
When distilled into its essence,
the ingredient we need to express
the concepts of cost and size
is
a product
\( \costInt{} \times \sizeInt{} \)
of a well-founded set \( \costInt{} \)
--- the cost set ---
and
a quasi-ordered set \( \sizeInt{} \)
--- the size set.
Intuitively, the cost tuples in \( \costInt{} \)
bound the number of rewriting steps
needed to reach normal forms,
which is in line with the aforementioned rewriting cost model.
Meanwhile,
the size tuples in \( \sizeInt{} \) are more general.
We can use integers, reals, and terms themselves as size.
Following the treatment in~\cite{kop:vale:21},
the construction of \cs{} products is done inductively
on the structure of types.
So we map each type \( \atype \) to a \cs{} product
\( \costInt{\atype} \times \sizeInt{\atype} \).
Hence, in this paper, our first-order term formalism follows a type discipline.

In order to extend the usability of our techniques,
we would like to
not only exhibit bounds to the runtime complexity
function but also determine sufficient conditions
for its feasibility, that is,
the existence of polynomial upper bounds.
In the eighties {Huet and Oppen}~\cite{huet:oppen:80}
conjectured that polynomial interpretations
are sufficient to evince feasibility,
which was disproved by
Lautemann~\cite{lautemann:88} in the same decade.
Indeed,
polynomial interpretations induce a double exponential
upper bound on the derivation length,
as shown by the seminal work of
{Hofbauer and Lautemann}~\cite{hof:lau:89}.
Feasibility can be recovered by imposing additional
conditions on interpretations.
To the best of our knowledge,
{Cichon and Lescanne}~\cite{cichon:lescanne:92}
were the first to propose
such conditions even though their setting is restricted
to number theoretic functions only.
Similar results are proved in~\cite{bonfante:cichon:marion:touzet:01},
where the authors provide rewriting
characterizations of complexity
classes using bounds for the interpretation of data constructors.
These same conditions appear in the higher-order setting,
see~\cite{baillot:lago:16,kop:vale:21}.
In the present paper,
we follow a similar approach to that
in~\cite{bonfante:cichon:marion:touzet:01} and show that
we can recover those classical results
by bounding size tuples in interpretations.

Tuple interpretations do not provide a complete termination
proof method:
there are terminating systems for which interpretations
cannot be found.
Consequently,
it does not induce a complete
complexity analysis framework either.
Notwithstanding,
it has the potential to be very powerful if we choose
the \cs{} sets wisely.
A second limitation is that the search for interpretations
is undecidable in general,
which is expected already in the
polynomial case~\cite{mitterwallner:aart:22}.
Undecidability never hindered computer scientists' efforts
on mechanizing difficult problems,
however.
Indeed, several proof search methods
have been developed over the years to find
interpretations
automatically~\cite{codish:gonopolskiy:ben-amram:amir:fuhs:giesl:11,
cherifa:lescanne:87,
contejan:marche:tomas:urbain:05,
hof:01,yamada:22
}.

\paragraph{Contribution.}

We provide a formal definition of \cs{} products
(Definition~\ref{def:cs-tuples})
and use it to interpret types in Definition~\ref{def:cost-size-sets}.
\csbegin{} products provide an interpretation domain for \cs{} tuple algebras,
Definition~\ref{def:cs-tuple-alg}.
In Lemmas~\ref{lemma:cs-sets-cs-tuple}~and~\ref{lemma:cs-int-is-sound}
we show the soundness of this approach.
In Definition~\ref{def:semantic-app}
we introduce a type-safe application
operator on \cs{} products
and prove its strong monotonicity,
an important ingredient to show the Compatibility
Theorem~\ref{thm:innermost-compatibility}.
We establish the termination of Toyama's system in
Example~\ref{ex:toyama-trs-is-it},
showing that Theorem~\ref{thm:innermost-compatibility}
correctly captures innermost termination in our setting.
We provide sufficient conditions so that feasible bounds on
innermost runtime complexity can be achieved
in Lemmas~\ref{lemma:add-linear-size-bounds}~and~\ref{lemma:int-bounds-imply-irc-bounds}.

\paragraph{Outline.}
In Section~\ref{sec:preliminaries},
we fix notation and recall basic notions of rewriting syntax,
basic terminology on the complexity of rewriting,
and review our notation for sets, orders, and functions.
In Section~\ref{sec:tuple-int},
we tailor tuple interpretations to the innermost setting
and prove the innermost version of the
compatibility theorem.
We proceed to establish complexity bounds to the innermost runtime
complexity in Section~\ref{sec:runtime-complexity}.
In Section~\ref{sec:hermes-description},
we present preliminary work on automation
techniques to find \cs{} tuple interpretations.
We conclude the paper in Section~\ref{sec:conc}.

\section{Preliminaries}\label{sec:preliminaries}

\paragraph*{TRSs and Innermost Rewriting.}

We consider simply typed first-order term rewriting systems in curried notation.
Fix a set \( \sortset \), whose elements are called \emph{sorts}.
The set \( \simpletypeset \) of \emph{types} is
generated by the grammar
\(
\simpletypeset \Coloneqq
\sortset
\mid
\sortset\arrtype\simpletypeset
\).
Each type is written as
\(\asort_1 \arrtype \cdots \arrtype \asort_m \arrtype \bsort \)
where all \(\asort_i\) and \(\bsort \) are sorts.
A \emph{signature} is a set \( \signature \) of symbols
together with an arity function \( \ar \)
which associates to each \( \afun \in \signature \) a type
\( \atype \in \simpletypeset \).
We call the triple \( (\sortset, \signature, \ar) \)
a \emph{syntax signature}.
For each sort \( \asort \),
we postulate a set \( \var_\asort \) of countably many variables
and assume that \( \var_\asort \cap \var_{\asort'} = \emptyset \)
if \(\asort \neq \asort'\).
Let \(\var \) denote \(\bigcup_\asort \var_\asort \) and assume that
 \(\signature \cap \var = \emptyset \).

The set \(\mathbb{T}\) of \emph{pre-terms} is generated by the grammar
\(\mathbb{T} \Coloneqq \signature \mid \var \mid (\mathbb{T} \app \mathbb{T})\).
The set \(\terms \) of \emph{terms} consists of pre-terms which can be typed as follows:
\begin{enumerate*}[label=(\roman*)] 
\item \(\afun \hasType \atype \)
    if \(\ar(\afun) = \atype \),
\item \(\avar \hasType \asort \)
    if \(\avar \in \var_\asort \), and
\item \((\aterm \app \bterm) \hasType \btype \)
    if \(\aterm \hasType \asort \arrtype \btype \)
    and
    \(\bterm \hasType \asort \).
\end{enumerate*}
Application of terms is left-associative,
so we write \(\aterm \app \bterm \app \cterm \) for
\(((\aterm \app \bterm) \app \cterm)\).
Let \(\vars{\aterm}\) be the set of variables
occurring in \(\aterm \).
A \emph{ground term} is a term \(\aterm \) such that \(\vars{\aterm}=\emptyset \).
A symbol \(\afun \in \signature \)
is called the \emph{head symbol}
of \(\aterm \) if
\(\aterm = \afun \app \aterm_1 \dots \aterm_k\).
A \emph{subterm} of \(\aterm\) is a term \(\bterm\) (we write \(\aterm \subtermR \bterm\)) such that
\begin{enumerate*}[label = (\roman*)]
\item \(\aterm = \bterm\), or
\item \(\bterm\) is a subterm of \(\aterm'\) or \(\aterm''\) when \(\aterm = \aterm' \app \aterm''\).
\end{enumerate*}
A \emph{proper subterm} of \(\aterm\) is a subterm of
\(\aterm\) which is not equal to \(\aterm\).
A \emph{substitution} \( \asub \) is a type-preserving map from variables to terms such that the set
\(\dom(\asub) = \{\avar \in \var \mid \asub(\avar) \neq \avar\}\) is finite.
Every substitution \(\asub\) extends to a type-preserving map from terms to terms, whose image on \(\aterm\) is written as \(\aterm\asub\), as follows:
\begin{enumerate*}[label = (\roman*)]
\item \(\afun \asub = \afun\),
\item \(\avar \asub = \asub(\avar)\), and
\item \((\aterm \app \bterm) \asub = (\aterm \asub) \app (\bterm \asub)\).
\end{enumerate*}

A relation \( \to \) on terms is \emph{monotonic}
if \(\aterm \to \aterm'\) implies
\(\bterm \app \aterm \to \bterm \app \aterm'\)
and \(\aterm \app \cterm \to \aterm' \app \cterm \)
for all terms \(\bterm \) and \(\cterm \) of appropriate types.
A \emph{rewrite rule} \( \ell \rulesArrow r \)
is a pair of terms of the same type
such that
\( \ell = \afun \app \ell_1 \dots \ell_k \)
and \( \vars{\ell} \supseteq \vars{r} \).
A \emph{term rewriting system} (TRS)
\( \rules \) is a set of rewrite rules.
The \emph{rewrite relation} \(\arrzR\) induced by \(\rules\)
is the smallest monotonic relation on terms such that
\(\ell \asub \arrzR r \asub\)
for all rules \(\ell \to r \in \rules\) and
substitutions \(\asub\).
A \emph{reducible expression} (redex) is a term of form \(\ell \asub\) for some rule \(\ell \rulesArrow r\) and substitution \(\asub\).
A term is in \emph{normal form}
if none of its subterms is a redex.
A TRS \(\rules\) is \emph{terminating} if no infinite rewrite sequence \(\aterm \arrzR \aterm' \arrzR \aterm'' \arrzR \cdots\) exists.

Every rewrite rule \(\ell \rulesArrow r\) \emph{defines} a symbol \(\afun\), namely, the head symbol of \(\ell\).
For each \(\afun \in \signature\), let \(\rules_\afun\) denote the set of rewrite rules that define \(\afun\) in \(\rules\).
A symbol \( \afun \in \signature \) is a
\emph{defined symbol} if
\(\rules_\afun \neq \emptyset\);
otherwise, \(\afun\) is called a \emph{constructor}.
Let \(\dfdS\) be the set of defined symbols and \(\ctrS\) the set of constructors.
So \( \signature = \dfdS \cup \ctrS \).
A \emph{data term} is a term of the form
\( \consFont{c} \app d_1 \app \dots \app d_k \)
where \( \consFont{c} \)
is a constructor and each \( d_i \) is a data term.
A \emph{basic term} is a term
of type \(\asort\) and
of form \( \afun \app d_1 \app \dots \app d_m \)
where \(\asort\) is a sort, \( \afun \) is a defined symbol and all
\( d_1, \dots, d_m \) are data terms.
We let \(\bTerms\) denote the set of all basic terms.

\begin{Exm}\label{ex:toy-trs}
    We fix \( \nat \) and \( \lst \) for the sorts of natural numbers and lists of natural numbers, respectively.
    In the below TRS,
    \( \zero \hasType \nat \),
    \( \suc \hasType \nat \arrtype \nat \),
    \( \nil \hasType \lst \) and
    \( \cons \hasType \nat \arrtype \lst \arrtype \lst \)
    are constructors while
    \( \add, \minus, \quot \hasType
    \nat \arrtype \nat \arrtype \nat \),
    \( \append \hasType \lst \arrtype \lst \arrtype \lst \),
    \( \sumList \hasType \lst \arrtype \nat \) and
    \( \rev \hasType \lst \arrtype \lst \)
    are defined symbols.
    \begin{align*}
        \add \app \avar \app \zero                                    & \to \avar
                                                                      & \sumList \app \nil                                                       & \to \zero \\
        \add \app \avar \app (\suc \app \bvar)
                                                                      & \to \suc \app (\add \app \avar \app \bvar)
                                                                      & \sumList \app (\cons \app \avar \app \aListVar)                          & \to
        \add \app (\sumList \app \aListVar) \app \avar                                                                                                       \\
        \append \app \nil \app \bListVar                              & \to
        \bListVar                                                     & \rev \app \nil                                                           & \to \nil  \\
        \append \app (\cons \app \avar \app \aListVar) \app \bListVar & \to
        \cons \app \avar \app (\append \app \aListVar \app \bListVar) &
        \rev \app (\cons \app x \app \aListVar)
                                                                      & \to \append \app (\rev \app \aListVar) \app (\cons \app \avar \app \nil)             \\
        \minus \app \avar \app \zero                                  & \arrz \avar                                                              &
        \quot \app \zero \app (\suc \app \bvar)                       & \arrz \zero                                                                          \\
        \minus \app \zero \app y                                      & \arrz \zero
                                                                      &
        \quot \app (\suc \app \avar) \app (\suc \app \bvar)
                                                                      & \arrz
        \suc \app ( \quot \app (\minus \app \avar \app \bvar ) \app (\suc \app \bvar))
        \\
        \minus \app (\suc \app \avar) \app (\suc \app \bvar)
                                                                      & \arrz
        \minus \app \avar \app \bvar
    \end{align*}
\end{Exm}

We restrict our attention to innermost rewriting:\ %
only redexes with no reducible proper subterms might be reduced.
More precisely, the \emph{innermost rewrite relation} \( \arrzR^i \)
induced by \( \rules \) is defined as follows:
\begin{enumerate}[label=(\roman*)]
    \item \( \ell\gamma \arrzR^i r\gamma \) if \( \ell \arrz r \in \rules \)
          and all proper subterms of \( \ell \gamma \) are in normal form, 
    \item \( \aterm \app \bterm \arrzR^i \aterm' \app \bterm \)
          if \( \aterm \arrzR^i \aterm' \), and
    \item \( \aterm \app \bterm \arrzR^i \aterm \app \bterm' \)
          if \( \bterm \arrzR^i \bterm' \).
\end{enumerate}
In this paper we only analyze innermost rewriting.
So
we write \(\arrz\) for \(\arrzR^i\) whenever no ambiguity arises.

\paragraph*{Derivation Height and Complexity.}

Given a well-founded and finitely branching relation \( \arrz \) on terms,
we write \( \aterm \xrightarrow{n} \bterm \)
if there is a sequence
\( \aterm = \aterm_0 \arrz \cdots \arrz \aterm_n = \bterm \)
of length \( n \).
The \emph{derivation height} \(\dht{\aterm, \arrz}\) of a term \( \aterm \)
with respect to \( \arrz \) is the length
of the longest \( \arrz \)-sequence of
starting with \( \aterm \),
i.e.,
\(
\dht{\aterm, \arrz} =
\max \{
    n \mid \exists \bterm \in \terms :
    \aterm \xrightarrow{n} \bterm
\}
\).
The \emph{absolute size} of a term \( \aterm \),
denoted by \( |\aterm| \),
is \( 1 \) if \( \aterm \) is a symbol in \( \signature \) or a variable,
and \( |\aterm_1| + |\aterm_2| \)
if \( \aterm = \aterm_1 \app \aterm_2 \).
In order to express various complexity notions in the rewriting setting,
we define the \emph{complexity function} as follows:
\(
\compFunc(n, \arrz, \mathcal{T}) =
\max \{
    \dht{\aterm, \arrz} \mid
    \aterm \in \mathcal{T} \text{ and } |\aterm| \leq n
\}
\).
Intuitively, \(\compFunc(n, \arrz, \mathcal{T})\) is the length of the longest \( \arrz \)-sequence starting with a term whose absolute size is at most \( n \) from \(\mathcal{T}\).
We summarize four particular instances in the following table:
\begin{align*}
                   &&             &\text{derivational}            &             &\text{runtime}\\
  \text{full}      && \dc(n)  ={} &\compFunc(n, \arrzR, \terms)   & \rc(n)  ={} &\compFunc(n, \arrzR, \bTerms)\\
  \text{innermost} && \idc(n) ={} &\compFunc(n, \arrzR^i, \terms) & \irc(n) ={} &\compFunc(n, \arrzR^i, \bTerms)
\end{align*}

\paragraph*{Ordered Sets and Monotonic Functions.}

A \emph{quasi-ordered set} \( (A, \sizeGe) \)
consists of a nonempty set \( A \)
and a quasi-order (reflexive and transitive)
\( \sizeGe \) on \( A \).
An \emph{extended well-founded set}
\( (A, \costGt, \costGe) \) is a nonempty set \( A \)
together with a well-founded order \(\costGt\)
and a quasi-order \(\costGe\) on \(A\)
such that \(\costGe\) is compatible with \( \costGt \),
i.e.,
\( x\costGt y \) implies \( x \costGe y \)
and \( x \costGt y \costGe z \) implies \( x > z \).
Below we refer to an extended well-founded set
simply as a \emph{well-founded set}.

Given quasi-ordered sets \( (A, \sizeGe) \) and \( (B, \sizeGe) \),
a function \( f : A \arrfunc B \) is said to be \emph{weakly monotonic} if \( x \sizeGe y \) implies \( f(x) \sizeGe f(y) \).
Let \( A \arrfuncwm B \)
denote the set of weakly monotonic functions from
\( A \) to \( B \).
The comparison operator \( \sizeGe \) on \(B\) induces pointwise comparison on \(A \arrfuncwm B\) as follows:
\( f \sizeGe g \) if
\( f(x) \sizeGe g(x) \) for all \( x \in A \).
This way \( (A \arrfuncwm B, \sizeGe) \) is also a quasi-ordered set.
Given well-founded sets \( (A, \costGt,\costGe) \) and \( (B,\costGt,\costGe) \),
a function \( f : A \arrfunc B \) is said to be \emph{strongly monotonic} if \( x \costGt y \) implies \( f(x) \costGt f(y) \) and \( x \costGe y \) implies \( f(x) \costGe f(y) \).

\section{Tuple Interpretations}\label{sec:tuple-int}

In this section,
we introduce the notion of tuple algebras in the
context of innermost rewriting.
We start by interpreting types as \cs{} products,
give interpretation of terms as \cs{} tuples,
and finally, prove the innermost version of the compatibility
theorem.

\subsection{Types as \cstitle{} Products}

We start by constructing a \cs{} denotational semantics to types in
\( \simpletypeset \).
The goal is to define a function \( \typeinterpret{\cdot} \)
that maps each type \( \atype \in \simpletypeset \)
to a well-founded set \( \typeinterpret{\atype} \),
the \cs{} interpretation of \( \atype \).

\begin{Def}[\cstitle{} Products]\label{def:cs-tuples}
  Given a well-founded set \( (\costInt{}, \costGt, \costGe) \),
  called the \emph{cost set},
  and a quasi-ordered set \( (\sizeInt{}, \sizeGe) \),
  called the \emph{size set},
  we call \( \costInt{} \times \sizeInt{} \) the \emph{\cs{} product} of
  \( (\costInt{}, \costGt, \costGe) \) and \( (\sizeInt{}, \sizeGe) \),
  and its elements \emph{\cs{} tuples}.
\end{Def}

Given a \cs{} product \( \costInt{} \times \sizeInt{} \),
the well-foundness of \( \costInt{} \) and quasi-ordering on \( \sizeInt{} \)
naturally induce an ordering structure on the cartesian product
\( \costInt{} \times \sizeInt{} \) as follows.

\begin{Def}[Product Order]\label{def:natural-prod-order}
    Let
    \(
        {(\costInt{}, \costGt, \costGe)}
        \times
        {(\sizeInt{}, \sizeGe)}
    \)
    be a \cs{} product.
    Then we define the relations
    \( {\cartGt}, {\cartGe} \) over
    \( \costInt{} \times \sizeInt{} \) as follows:
    for all \( \pair{x,y} \) and \( \pair{x',y'} \) in
    \( \costInt{} \times \sizeInt{} \),
    \begin{enumerate}[label=(\roman*)]
        \item \( \pair{x,y} \cartGt \pair{x',y'} \) if \( x \costGt x' \) and \( y \sizeGe y' \), and
        \item \( \pair{x,y} \cartGe \pair{x',y'} \) if \( x \costGe x' \) and \( y \sizeGe y' \).
    \end{enumerate}
\end{Def}
Next,
we show that \cs{} products ordered as above form a well-founded set.
\begin{Lem}\label{lemma:natural-prod-is-wf}
    The triple \( (\costInt{} \times \sizeInt{}, \cartGt, \cartGe) \)
    is a well-founded set.
\end{Lem}
\begin{proof}
  It follows immediately from Definition~\ref{def:cs-tuples}
  that \( \cartGt, \cartGe \) are transitive
  and \( \cartGe \) is reflexive.
  To prove that \( \cartGt \) is well-founded,
  note that the existence of
  \( \pair{x_1, y_1} \cartGt \pair{x_2, y_2} \cartGt \cdots \)
  would imply \( x_1 \costGt x_2 \costGt \cdots \)
  which cannot be the case since \( \costGt \) is well-founded.

  We still need to check that \( \cartGe \) is compatible with \( \cartGt \).
  \begin{itemize}
  \item Suppose \( \pair{x,y} \cartGt \pair{x', y'} \).
    Since \( x \costGt x' \) implies \( x \costGe x' \),
    we have \( \pair{x,y} \cartGe \pair{x',y'} \).
  \item Suppose \( \pair{x,y} \cartGt \pair{x',y'} \cartGe \pair{x'', y''} \).
    Since \( x \costGt x' \costGe x'' \) implies
    \( x \costGt x'' \) and \( \sizeGe \) is transitive,
    we have \( \pair{x,y} \cartGt \pair{x'', y''} \).
  \end{itemize}
\end{proof}

Now we interpret types as a particular kind of \cs{} products.
\begin{Def}[Interpretation of Types]\label{def:cost-size-sets}
    Let \( \sortset \) denote the set of sorts.
    An \emph{interpretation key} \( \intKey \) for \( \sortset \) maps each sort
    \( \asort \) to a quasi-ordered set \( (\intKey(\asort), \sizeGe) \) with a minimum.
    For each type \( \atype \in \simpletypeset \),
    we define the \cs{} interpretation of \( \atype \)
    as the product \( \typeinterpret{\atype} = \costInt{\atype} \times \sizeInt{\atype} \)
    with
    \begin{align*}
      \costInt{\atype}                  & = \Nat \times \costFInt{\atype}                & \\
      \costFInt{\asort}                 & = \unit                                        & \sizeInt{\asort}                  & = \intKey(\asort)\\
      \costFInt{\asort \arrtype \btype} & = \sizeInt{\asort} \arrfuncwm \costInt{\btype} & \sizeInt{\asort \arrtype \btype}  & = \sizeInt{\asort} \arrfuncwm \sizeInt{\btype}
    \end{align*}
    where \( \unit = \{ \unitE \} \) is quasi-ordered by \( \costGe \) with \( \unitE \costGe \unitE \).
    All \( \costFInt{\asort \arrtype \btype} \) and \( \sizeInt{\asort \arrtype \btype} \) are ordered by pointwise comparison.
    The set \( \costInt{\atype} \) is ordered as follows:
    \( (n, f) \costGt (m, g) \)
        if \( n > m \) and \( f \costGe g \), and
    \( (n, f) \costGe (m, g) \)
        if \( n \costGe m \)  and \( f \costGe g \).
    This definition requires that all \( (\costInt{\atype}, \costGe) \) and \( (\sizeInt{\atype}, \sizeGe) \) are quasi-ordered sets,
    which is guaranteed by the following lemma.
\end{Def}
\begin{Lem}\label{lemma:cs-sets-cs-tuple}
    For any type \( \atype \),
    \( (\costInt{\atype}, \costGt, \costGe) \) is a well-founded set
    and \( (\sizeInt{\atype}, \sizeGe) \) is a quasi-ordered set with a minimum.
    Therefore, \( \typeinterpret{\atype} \)
    is a \cs{} product.
\end{Lem}
\begin{proof}
  When \( \atype \) is a sort, \( \costInt{\atype} = \Nat \times \unit \cong \Nat \) and \( \sizeInt{\atype} = \intKey(\atype) \), so the statement is trivially true.
  When \( \atype = \asort \arrtype \btype \), we have \( \costInt{\atype} = \Nat \times \costFInt{\asort \arrtype \btype} \), \( \costFInt{\asort \arrtype \btype} = \intKey(\asort) \arrfuncwm \costInt{\btype} \) and \( \sizeInt{\atype} = \intKey(\asort) \arrfuncwm \sizeInt{\btype} \).
  By induction, \( (\costInt{\btype}, \costGe) \) and \( (\sizeInt{\btype}, \sizeGe) \) are quasi-ordered sets.
  So are \( (\costFInt{\asort \arrtype \btype}, \costGe) \) and \( (\sizeInt{\atype}, \sizeGe) \), which are ordered by pointwise comparison.
  By Lemma~\ref{lemma:natural-prod-is-wf}, \( (\costInt{\atype}, \costGt, \costGe) \) is a well-founded set.
  One minimum of \( (\sizeInt{\atype}, \sizeGe) \) is the constant function \( \fatlambda x. \bot \) where \( \bot \) is a minimum of \( (\sizeInt{\btype}, \sizeGe) \).
\end{proof}

The cost component \( \costInt{\atype} \)
of \( \typeinterpret{\atype} \) holds information about the cost of
reducing a term of type \( \atype \) to its normal form.
It has two parts: one is numeric; the other is functional.
The functional part \( \costFInt{\atype} \) degenerates to \( \unit \) when \( \atype \) is just a sort
and is indeed a functional space when
\( \atype = \asort \arrtype \btype \) is a function type.
In the latter case, \( \costFInt{\atype} = \sizeInt{\asort} \arrfuncwm \costInt{\btype} \) consists of
weakly monotonic functions with domain \( \sizeInt{\asort} \),
the size component of \( \typeinterpret{\asort} \).
This is very much in line with the standard
complexity notion based on Turing Machines
in which time complexity is
parametrized by the input's size.

We need a concrete interpretation key
in order to use Definition~\ref{def:cost-size-sets}
to interpret types.
In our examples,
a particular kind of interpretation key
maps each sort \( \asort \) to
size sets of the form
\( (\Nat^{\typecount{\asort}}, \sizeGe) \),
with \( \typecount{\asort} \geq 1 \),
and are ordered as follows:
\( {\pair{x_1, \dots, x_{\typecount{\asort}}}}
\sizeGe {\pair{y_1, \dots, y_{\typecount{\asort}}}} \)
if \( x_i \geq y_i \) 
for all \( i \).
This class of interpretation key
is used unless stated otherwise.
We take a semantic approach
(cf.~\cite{kop:vale:21})
to determine the number \( \typecount{\asort} \)
for each sort~\( \asort \).
For instance
\( \nat \) is the sort of natural numbers in unary format,
so a number \( n \in \Nat \) is represented as
the data term \( \suc \app {(\ldots {(\suc \app \zero)})} \),
that is,
\( n \) successive applications of \( \suc \) to \( \zero \).
With that in mind
the number of occurrences of \( \suc \) in such terms
is a reasonable measure for their size,
so we let \( \typecount{\nat} = 1 \).
A second example is that of \( \lst \).
To characterize the size of a list
we may need information about the individual elements
in addition to the length of the list.
So we keep track of the length as well as
the maximum size of their elements.
This way \( \typecount{\lst} = 2 \).
In Example~\ref{ex:int-constructors}
we interpret \( \nat, \lst \) constructors following
this intuition.

\begin{Def}\label{def:cost-size-fct}
  \csbegin{} tuples in \( \typeinterpret{\atype} \) are written as \( \pair{(n, f^\cost), f^\size} \)
  where \( n \in \Nat \), \( f^\cost \in \costFInt{\atype} \), and \( f^\size \in \sizeInt{\atype} \).
  When \( \atype \) is a function type,
  we refer to \( f^\cost \) as the \emph{cost function}
  and \( f^\size \) as the \emph{size function}.
\end{Def}

In order to define the interpretation of terms (Definition~\ref{def:term-int}),
we need a notion of application for \cs{} tuples.
More precisely,
given \( \tuple{f} \in \typeinterpret{\asort \arrtype \btype} \)
and \( \tuple{x} \in \typeinterpret{\asort} \),
our goal is to define
\( \tuple{f} \semApp \tuple{x} \in \typeinterpret{\btype} \),
the application of \( \tuple{f} \) to \( \tuple{x} \).
Let us illustrate how such an application should work with a concrete example.
Consider the function
\( \append \hasType \lst \arrtype \lst \arrtype \lst \)
from Example~\ref{ex:toy-trs}.
It takes two lists \( \aListVar \) and \( \bListVar \) as input.
The intended \cs{} denotational semantics for \( \append \)
is a tuple
\(
    \tuple{f} =
    \pair{(n, f^\cost), f^\size} \in
    \typeinterpret{\lst \arrtype \lst \arrtype \lst}
\), where
\begin{align*}
  n      &\in \Nat\text{,}\\
  f^\cost &\in \overbrace{\sizeInt{\lst}}^{\text{size of } \aListVar}
              \arrfuncwm (\Nat \times
                  (
                      \overbrace{\sizeInt{\lst}}^{\text{size of } \bListVar }
                      \arrfuncwm (\Nat \times \unit)
                  )
              )\text{, and}\\
  f^\size &\in \overbrace{\sizeInt{\lst}}^{\text{size of } \aListVar} \arrfuncwm ( \overbrace{\sizeInt{\lst}}^{\text{size of } \bListVar } \arrfuncwm \sizeInt{\lst} )\text{.}
\end{align*}
For the first list \( \aListVar \), take a \cs{} tuple \( \tuple{x} = \pair{(m, \unitE), x^\size} \) from \( \typeinterpret{\lst} \).
We apply \( f^\cost \) and \( f^\size \) to \( x^\size \),
and get \( f^\cost(x^\size) = (k, h) \in \Nat \times (\sizeInt{\lst} \arrfuncwm (\Nat \times \unit)) \) and \( f^\size(x^\size) \in \sizeInt{\lst} \arrfuncwm \sizeInt{\lst} \), respectively.
Then we sum the numeric parts and collect all the data in the new \cs{} tuple \( \pair{(n + m + k, h), f^\size(x^\size)} \).
This process is summarized in the following definition.

\begin{Def}[Semantic Application]\label{def:semantic-app}
    Given \( \tuple{f} = \pair{(n, f^\cost), f^\size} \in \typeinterpret{\asort \arrtype \btype} \)
    and \( \tuple{x} = \pair{(m, \unitE), x^\size} \in \typeinterpret{\asort} \),
    the \emph{semantic application} of \( \tuple{f} \) to \( \tuple{x} \),
    denoted by \( \tuple{f} \semApp \tuple{x} \),
    is \( \pair{(n + m + k, h), f^\size(x^\size)} \) where
    \( f^\cost(x^\size) = (k, h) \).
\end{Def}
Semantic application is left-associative,
so \( \tuple{f} \semApp \tuple{g} \semApp \tuple{h} \)
stands for \( (\tuple f \semApp \tuple g) \semApp \tuple h \).
This definition conforms to the types,
which is stated in the following lemma.

\begin{Lem}\label{lemma:semApp-type-sound}
    If \( \tuple{f} \in \typeinterpret{\asort \arrtype \btype} \)
    and \( \tuple{x} \in \typeinterpret{\asort} \),
    then \( \tuple{f} \semApp \tuple{x} \in \typeinterpret{\btype} \).
\end{Lem}

\begin{Rem}
  Because \( \Nat \times \unit \) is order-isomorphic to \( \Nat \),
  we identify \( \Nat \times \unit \) with \( \Nat \) and \( (m, \unitE) \) with \( m \)
  unless otherwise stated.
  So we write \( \pair{m, x^\size} \) for \cs{} tuples in \( \typeinterpret{\asort} \) where \( \asort \) is a sort.
\end{Rem}

\subsection{\cstitle{} Tuple Algebras}\label{subsec:cs-tuple-algebras}

An interpretation of a syntax signature
\( (\sortset, \signature, \ar) \)
interprets the types in \( \simpletypeset \)
and each \( \afun \hasType \atype \in \signature \)
to an element of \( \typeinterpret{\atype} \).
This is formally stated in the definition below.
\begin{Def}\label{def:cs-tuple-alg}
    A \emph{\cs{} tuple algebra}
    \( (\typeinterpret{\cdot}, \funcinterpret{}) \) over a syntax signature
    \( (\sortset, \signature, \ar) \)
    consists of:
    \begin{enumerate}[label=(\roman*)]
    \item a family of \cs{} products
      \( {\{\typeinterpret{\atype}\}}_{\atype \in \simpletypeset} \), and
    \item an interpretation function
      \( \funcinterpret{}
      : \signature \arrfunc
      \biguplus_{\atype} \typeinterpret{\atype}
      \)
      that associates to each \( \afun \hasType \atype \)
      an element
       \( \funcinterpret{\afun} \in \typeinterpret{\atype} \).
    \end{enumerate}
\end{Def}

We extend the notion of interpretation to terms,
where we use a valuation to map variables of type
\( \asort \) to elements of \( \typeinterpret{\asort} \).
With innermost rewriting we assume
that variables have no cost.

\begin{Def}\label{def:term-int}
  Fix a \cs{} tuple algebra \( (\typeinterpret{\cdot}, \funcinterpret{}) \).
  A \emph{valuation} \( \alpha : \var \arrfunc \biguplus_{\atype} \typeinterpret{\atype} \) is a function
  which maps each variable \( \avar \hasType \asort \) to a zero-cost tuple \( \pair{0, x^\size} \in \typeinterpret{\asort} \).
  The interpretation of a term \( \aterm \) under the valuation \( \alpha \),
  denoted by \( \ainterpret{\aterm} \), is defined as follows:
    \begin{align*}
        \ainterpret{\afun} &= \funcinterpret{\afun}
        &
        \ainterpret{\avar} &= \alpha(\avar)
        &
        \ainterpret{\aterm \app \bterm}
        &= \ainterpret{\aterm} \semApp \ainterpret{\bterm}
    \end{align*}
\end{Def}

We write \( \interpret{\aterm} \) instead of \( \ainterpret{\aterm} \)
whenever \( \alpha \) and \( \funcinterpret{} \) are universally quantified
or clear from the context.
In both cases we may write \( \interpret{\avar} = x \) instead of
\( \ainterpret{\avar} = \alpha(\avar) \).
As a corollary of Lemma~\ref{lemma:semApp-type-sound},
the interpretation of terms conforms with types.
\begin{Lem}\label{lemma:cs-int-is-sound}
    If \( \aterm \hasType \atype \) then
    \( \interpret{\aterm} \in \typeinterpret{\atype} \).
\end{Lem}

Let \( \atype \) be \( \asort_1 \arrtype \dots \arrtype \asort_m \arrtype \bsort \)
where all \(\asort_i\) and \(\bsort \) are sorts.
Elements of \( \costInt{\atype} \) can be written as
\begin{equation}\label{eq:cost-tuple-full-form}
  \begin{split}
    &          (e_0, \fatlambda x_1.\\
    &\hphantom{(e_0, \fatlambda x_1.}(e_1, \fatlambda x_2.\\
    &\hphantom{(e_0, \fatlambda x_1.(e_1, \fatlambda x_2.}\ldots\\
    &\hphantom{(e_0, \fatlambda x_1.(e_1, \fatlambda x_2.\ldots}(e_{m-1}, \fatlambda x_m.\\
    &\hphantom{(e_0, \fatlambda x_1.(e_1, \fatlambda x_2.\ldots(e_{m-1}, \fatlambda x_m.}(e_m, \unitE))\ldots)).
  \end{split}
\end{equation}
When \( e_0 = e_1 = \cdots = e_{m - 1} = 0 \),
we write \( \costWrapper{\fatlambda x_1 \dots x_m. e_m } \)
as a shorthand.

\begin{Exm}\label{ex:int-constructors}
  Let \( \sizeInt{\nat} \) and \( \sizeInt{\lst} \) be \( \Nat \) and \( \Nat \times \Nat \), respectively.
  Recall that the size of a natural number is the number of occurrences of \( \suc \),
  and the size of a list is a pair \( q = (q_\leng, q_\mmax) \) where \( q_\leng \) is the length and \( q_\mmax \) is the maximum size of the elements.
  We interpret the constructors as follows:
    \begin{align*}
        \funcinterpret{\zero} & =
        \pair{0, 0}
        &
        \funcinterpret{\suc}  & =
        \pair{\costWrapper{\fatlambda x. 0}, \fatlambda x. x + 1}   \\
        \funcinterpret{\nil}  & = \pair{0,(0,0)}
        &
        \funcinterpret{\cons} & = \pair{
            \costWrapper{\fatlambda x q. 0},
            \fatlambda x q. (q_\leng + 1, \max(x, q_\mmax)) }
    \end{align*}
    Both \( \zero \) and \( \nil \) have no cost because they are constructors without a function type.
    With innermost rewriting, constructors with a function type, such as \( \suc \) and \( \cons \),
    have \( e_0 = \cdots = e_m = 0 \) for cost, of form~(\ref{eq:cost-tuple-full-form}).
\end{Exm}

\begin{Rem}\label{rmk:vars-no-cost}
    In Definition~\ref{def:term-int} we require that
    valuations interpret variables as zero-cost tuples.
    This is an important but subtle requirement
    that only works when reductions are innermost.
    Indeed,
    if reduction is unrestricted
    we can instantiate variables on the left-hand side of rules
    to terms containing redexes
    for which the cost should be accounted.
    Hence, not accounting for the cost of variables in full rewriting
    would lead to unsound analysis.
    Additionally,
    zero-cost tuples allow us to prove the innermost termination
    of the TRS \( \rules \) in Example~\ref{ex:toyama-trs-is-it},
    which is non-terminating in full rewriting.
\end{Rem}

\subsection{Compatibility Theorem}\label{sub:sec:innermost-compt-theorem}

Roughly, the compatibility theorem
(Theorem~\ref{thm:innermost-compatibility})
states that if \( \rules \)
is compatible with a tuple algebra \( \aDom \),
then the innermost rewrite relation \( \arrzR^i \)
is embedded in the well-founded order on \cs{} products.
The next two lemmas are technical results needed in order to prove it.
Lemma~\ref{lemma:sbst:lemma} states that interpretations are closed under substitution
and Lemma~\ref{lemma:sem-app-is-sm} provides strong monotonicity to semantic application.

\begin{Def}\label{def:subst-lift}
  Fix a \cs{} tuple algebra \( (\typeinterpret{\cdot}, \funcinterpret{}) \).
  A substitution \( \asub \) is \emph{zero-cost} under valuation \( \alpha \) if \( \ainterpret{\asub(\avar)} \) is a zero-cost tuple for each variable \( \avar \).
  Given a valuation \( \alpha \) and a zero-cost substitution \( \asub \),
  the function \( \alpha^\asub = \ainterpret{\cdot} \circ \asub = \ainterpret{\asub(\cdot)} \) is thus a valuation.
\end{Def}
\begin{Lem}[Substitution]\label{lemma:sbst:lemma}
  If \( \asub \) is a zero-cost substitution under valuation \( \alpha \), \( \ainterpret{\aterm\asub} = \interpret{\aterm}_{\alpha^\asub}^{\funcinterpret{}} \) for any term \( \aterm \).
\end{Lem}

\begin{Lem}\label{lemma:sem-app-is-sm}
    The application functional
    \( \semAppFunc(\tuple{f}, \tuple{x}) = \tuple{f} \semApp \tuple{x} \) is strongly monotonic on both arguments.
\end{Lem}
\begin{proof}
    We need to prove
    \begin{enumerate*}[label=(\roman*)]
    \item if \( \tuple{f} \cartGt \tuple{g} \) and \( \tuple{x} \cartGe \tuple{y} \),
    then \( \semAppFunc(\tuple f, \tuple x) \cartGt \semAppFunc(\tuple{g}, \tuple{y}) \);

    \item if \( \tuple{f} \cartGe \tuple{g} \) and \( \tuple{x} \cartGt \tuple{y} \),
    then \( \semAppFunc(\tuple f, \tuple{x}) \cartGt \semAppFunc(\tuple{g}, \tuple{y}) \);
    \item if \( \tuple{f} \cartGe \tuple{g} \) and \( \tuple{x} \cartGe \tuple{y} \),
    then \( \semAppFunc(\tuple f, \tuple{x}) \cartGe \semAppFunc(\tuple{g}, \tuple{y}) \).
    \end{enumerate*}
    Consider \cs{} tuples \( \tuple{f}, \tuple{g} \in \typeinterpret{\asort \arrtype \btype} \) and
    \( \tuple{x}, \tuple{y} \in \typeinterpret{\asort} \).
    Let \( \tuple{f} = \pair{(n, f^\cost), f^\size} \), \( \tuple{g} = \pair{(m, g^\cost), g^\size} \),
    \( \tuple{x} = \pair{x^\cost, x^\size} \), and \( \tuple{y} = \pair{y^\cost, y^\size} \).
    We proceed to show (i) and observe that (ii) and
    (iii) follow similar reasoning.
    Indeed,
    if \( \tuple{f} \cartGt \tuple g \) and \( \tuple{x} \cartGe \tuple{y} \) we have that
    \( n \costGt m, f^\cost \costGe g^\cost \), \(f^\size \sizeGe g^\size \),
    \( x^\cost \costGe y^\cost \), and \(x^\size \sizeGe y^\size \).
    Hence,
    by letting \( f^\cost(x^\size) = (k,h) \) and \( g^\cost(y^\size) = (k', h') \),
    we get:
    \[
        \semAppFunc(\tuple{f}, \tuple{x})
        =
        \pair{(n, f^\cost), f^\size}
        \semApp
        \pair{x^\cost, x^\size}
        =
        \pair{(n + x^\cost + k, h), f^\size(x^\size)}
        \costGt
        \pair{(m + y^\cost + k', h'), g^\size(y^\size)}
        =
        \semAppFunc(\tuple g, \tuple y)
    \]
\end{proof}

\begin{Def}\label{def:comp-trs-tuple-alg}
    A TRS \( \rules \) is said to be \emph{compatible} with a \cs{} tuple algebra \( (\typeinterpret{\cdot}, \funcinterpret{}) \) if
    \( \ainterpret{\ell} \cartGt \ainterpret{r} \) for all rules \( \ell \rulesArrow r \in \rules \) and valuations \( \alpha \).
\end{Def}

\begin{Thm}[Compatibility]\label{thm:innermost-compatibility}
    Let \( \rules \) be a TRS compatible with a \cs{} tuple algebra \( (\typeinterpret{\cdot}, \funcinterpret{}) \).
    Then, for any pair of terms \( \aterm \) and \( \bterm \), whenever \( \aterm \arrzR^i \bterm \)
    we have \( \ainterpret{\aterm} \cartGt \ainterpret{\bterm} \).
\end{Thm}
\begin{proof}
    We proceed by induction on \( \arrzR^i \).
    For the base case, \( \aterm \arrzR^i \bterm \) by \( \ell \asub \arrz r\asub \)
    and all subterms of \( \ell \asub \) are in \( \arrzR \) normal form.
    Therefore, since \( \ainterpret{\ell} \cartGt \ainterpret{r} \) by hypothesis, Lemma~\ref{lemma:sbst:lemma}
    gives us that \( \ainterpret{\ell\asub} \cartGt \ainterpret{r\asub} \).

    In the inductive step we use Lemma~\ref{lemma:sem-app-is-sm} combined with the (IH) as follows.
    Suppose \( \aterm \arrzR^i \bterm \) by \( \aterm = \aterm' \app \!\! \cterm \arrzR^i \aterm'' \cterm \)
    with \( \aterm' \arrzR^i \aterm'' \).
    Hence,
    \(
    \ainterpret{\aterm' \!\! \app \cterm} =
    \ainterpret{\aterm'} \semApp \ainterpret{\cterm} =
    \semAppFunc(\ainterpret{\aterm'}, \ainterpret{\cterm})
    \),
    henceforth the induction hypothesis gives
    \( \ainterpret{\aterm'} \cartGt \ainterpret{\aterm''} \),
    which combined with Lemma~\ref{lemma:sem-app-is-sm} implies
    \(
    \ainterpret{\aterm} =
    \semAppFunc(\ainterpret{\aterm'},\ainterpret{\cterm})
    \cartGt
    \semAppFunc(\ainterpret{\aterm''}, \ainterpret{\cterm})
    =
    \ainterpret{\bterm}
    \).
    When \( \aterm \arrzR^i \bterm \) with
    \(
    \aterm
    =
    \aterm' \app \cterm \arrzR^i \aterm' \app \cterm' \)
    the proof is analogous.
\end{proof}

\begin{Exm}\label{ex:toyama-trs-is-it}
    Let \( \num{0}, \num{1} \hasType \asort \),
    \( \bfun \hasType \asort \arrtype \asort \arrtype \asort \),
    and \( \afun \hasType \asort \arrtype \asort \arrtype \asort \arrtype \asort \).
    The rewrite system
    introduced by Toyama~\cite{toyama:87}
    and
    defined by
    \(
        \rules = \{
        \bfun \app \avar \app \bvar \arrz \avar,\,
        \bfun \app \avar \app \bvar \arrz \bvar,\,
        \afun \app \num{0} \app \num{1} \app \cvar \arrz \afun \app \cvar \app \cvar \app \cvar
    \}
    \)
    was given to show that termination is not modular for disjoint unions of TRSs.
    Indeed, it admits the infinite rewriting sequence
    \(
    \afun \app \num{0} \app \num{1}
    \app
    (\bfun \app \num 0 \app \num 1)
    \arrzR
    \afun \app (\bfun \app \num 0 \app \num 1)
    \app
    (\bfun \app \num 0 \app \num 1) \app
    (\bfun \app \num 0 \app \num 1) \arrzR^+
    \afun \app \num{0} \app \num{1}
    \app
    (\bfun \app \num 0 \app \num 1)
    \).
    However, the innermost relation \( \arrzR^i \) is terminating.
    In order to prove it,
    we introduce a non-numeric notion of size.
    Let \( \intKey(\asort) = \mathcal{P}(\terms) \), i.e.,
    the set of all subsets of \( \terms \).
    This set is partially ordered by set inclusion,
    so \( x \sizeGe y \) iff \( x \supseteq y \), which is a quasi-order.
    Consider the following interpretation:
    \begin{align*}
        \funcinterpret{\num{0}} &= \pair{0, \{\num{0}\}} &
        \funcinterpret{\num{1}} &= \pair{0, \{\num{1}\}} &
        \funcinterpret{\bfun}   &= \pair{\costWrapper{\fatlambda x y. 1}, \fatlambda x y. x \cup y} &
        \funcinterpret{\afun}   &= \pair{\costWrapper{\fatlambda xyz. H(x,y)},
        \fatlambda xyz. \emptyset},
    \end{align*}
    where \( H \) is a helper function defined by
    \( H(x,y) = \mathtt{if} \ x \sizeGe \{\num{0}\} \wedge y \sizeGe \{\num{1}\}
    \ \mathtt{then} \ 1 \
    \mathtt{else} \ 0 \).
    Notice that \( H \) is weakly monotonic
    and all terms in normal form are interpreted as sets of size \( \leq \) 1.
    Checking compatibility is straightforward:
    \( \interpret{\bfun \app \avar \app \bvar} =
    \pair{1, \avar \cup \bvar} \cartGt \pair{0, x} = \interpret{\avar} \) and
    \( \interpret{\bfun \app \avar \app \bvar} =
    \pair{1, \avar \cup \bvar} \cartGt \pair{0, \bvar} = \interpret{\bvar} \);
    and
    \( \interpret{\afun \app \num{0} \app \num{1} \app \cvar} =
    \pair{1, \emptyset} \cartGt
    \pair{0, \emptyset} = \interpret{\afun \app \cvar \app \cvar \app \cvar} \),
    because any instantiation of \( z \) is necessarily in normal form,
    so it cannot include both \( \num{0} \) and \( \num{1} \).
\end{Exm}

This example, albeit artificial, is interesting from a termination point of view.
It shows that tuple interpretations can be used
to deal with rewrite systems that only terminate
via the innermost strategy.

\section{Polynomial Bounds for Innermost Runtime Complexity}\label{sec:runtime-complexity}

In this section,
we study the applications of tuple interpretations
to complexity analysis of compatible TRSs,
i.e., rewriting systems that admit an interpretation
in a tuple algebra \( (\typeinterpret{\cdot}, \funcinterpret{}) \).
Even though cost and size
are split in our setting,
they are intertwined concepts
(in a sense we make precise in this section)
that constitute what we intuitively
call ``complexity'' of a TRS\@.

\subsection{Additive Tuple Interpretations}

In order to establish upper bounds to \(\irc(n)\),
it suffices to give upper bounds to the cost component
\( \interpret{\aterm}^\cost \) of
all terms \(\aterm \)
where \( |\aterm| \leq n \).
Furthermore,
since basic terms are of the form
\( \afun \app d_1 \dots d_m \),
the size of data terms plays an important role in our analysis.
In what follows,
we use the default choice for interpretation key
when interpreting types; that is,
\( \intKey(\asort) = \NatSizeSet{\asort} \),
with \( \typecount{\asort} \geq 1 \) for each
\( \asort \in \sortset \).

Given
\( \atype =
    \asort_1 \arrtype
    \dots
    \arrtype
    \asort_m \arrtype
\bsort \),
the size component
of \( \typeinterpret{\atype} \)
is
\( \sizeInt{\atype} =
\NatSizeSet{\asort_1} \arrfuncwm
\dots
\arrfuncwm
\NatSizeSet{\asort_m} \arrfuncwm
\NatSizeSet{\bsort} \).
Size functions \( f^\size \in \sizeInt{\atype} \)
when fully applied
can be written in terms of functional components.
Hence,
\( f^\size(x_1, \dots, x_m) =
\pair{
    f^\size_1(x_1, \dots, x_m),
    \dots,
    f^\size_{\typecount{\bsort}}(x_1, \dots, x_m)
} \).

\begin{Def}\label{def:size-int-add-lin-bounded}
    Let \( \atype \) be a type and \( f^\size \in \sizeInt{\atype} \).
    The size function \( f^\size \) is \emph{linearly bounded}
    if each one of its component functions
    \( f^\size_1, \dots, f^\size_{\typecount{\bsort}} \)
    is upper-bounded by a positive linear polynomial, i.e.,
    there is a positive constant \(a \in \Nat\) such that
    for all \( 1 \leq l \leq m \),
    \( f^\size_l(x_1, \dots, x_m) \leq
    a(1 + \sum_{i = 1}^m\sum_{j = 1}^{\typecount{\asort_i}}
    x_{ij})\).
    Analogously,
    we say \( f^\size \) is \emph{additive}
    if there is a constant \(a \in \Nat\) such that
    \(
        \sum_{l = 1}^{\typecount{\bsort}}
        f^\size_l(x_1, \dots, x_m)
        \leq
        a + \sum_{i = 1}^m\sum_{j = 1}^{\typecount{\asort_i}}
            x_{ij}
    \).
\end{Def}

Notice that by this definition
linearly bounded (or additive) size functions
are not required to be linear (or additive)
but to be upper-bounded by a linear (additive)
function.
So this permits us to use for instance
\( \min(x, 2 y) \),
whereas \( x y \) cannot be used.
Size interpretations do not necessarily bound the absolute size of data terms.
For instance,
we may interpret a data constructor
\( \consFont{c} \hasType \asort \arrtype \bsort \)
with \( \funcinterpret{\consFont{c}}^\size = \fatlambda x. \floor{x / 2} \)
which would give us \( |d| \geq \interpret{d}^s \).
This is especially useful when dealing with sublinear interpretations.

The next lemma ensures that by interpreting constructors additively,
the size interpretation of data terms is proportional to their absolute size:

\begin{Lem}\label{lemma:add-linear-size-bounds}
    Let \( \rules \) be a TRS compatible with a \cs{} tuple algebra
    \( (\typeinterpret{\cdot}, \funcinterpret{}) \).
    \begin{enumerate}[label = (\roman*)]
        \item
        Assume \( \funcinterpret{\consFont{c}}^\size \)
        is additive for all data constructors
        \( \consFont{c} \),
        then for all data terms \( d \):
        if \( |d| \leq n \),
        then there exists a constant \( b > 0 \) such that \( \interpret{d}^\size_l \leq bn \),
        for each size-component \( \interpret{d}^\size_l \) of \( \interpret{d} \).

        \item Assume \( \funcinterpret{\consFont{c}}^\size \)
        is linearly bounded for all data constructors \( \consFont{c} \),
        then for all data terms \( d \):
        if \( |d| \leq n \),
        then there exists a constant \( b > 0 \) such that \( \interpret{d}^\size_l \leq 2^{b n} \),
        for each size-component \( \interpret{d}^\size_l \) of \( \interpret{d} \).
    \end{enumerate}
\end{Lem}
The bound in (ii) is sharp.
Indeed,
define (when interpreting \( \rules_{\add}\)):
\( \funcinterpret{\zero} = \pair{0, 1} \),
\( \funcinterpret{\suc} = \pair{
    \costWrapper{\fatlambda x. 0},
    \fatlambda x. 2x + 1
    }
 \), and
\( \funcinterpret{\add} = \pair{
    \costWrapper{\fatlambda x y. y + 1},
    \fatlambda x y. x + y
}\).
In this case,
for a data term \( \num{n} = \suc^n(0) \)
its size interpretation is exactly
\( \interpret{\num{n}}^\size = 2^n + n \leq 2^{|\num{n}|} \).
However,
whereas this choice is compatible with \( \rules_{\add} \),
and hence proving its termination,
it induces an exponential overhead on
\( \irc[\add] \),
which is linearly bounded (see Example~\ref{ex:int-def-symbols}).
Such a huge overestimation is not desirable in a complexity analysis setting.
This behavior sets a strict upper-bound
to the interpretation of data constructors;
namely,
we seek to bound the size interpretations of constructors additively.
It is easy to show that size components for \( \nat \) and \( \lst \)
in Example~\ref{ex:int-constructors} are additive.

\begin{Def}\label{def:additive-int}
    We say an interpretation \(\funcinterpret{}\) is additive
    if for each \( \consFont{c} \in \ctrS \),
    \( \funcinterpret{\consFont{c}}^\size \) is additive.
\end{Def}

\subsection{Cost-Bounded Tuple Interpretations}

In what follows,
we consider rewriting systems with
additive interpretations.

\begin{Def}\label{def:cost-tuple-bounds}
    Let \( \atype \) be a type and \( \tuple{f}^\cost \in \costInt{\atype} \).
    We say \( \tuple{f}^\cost \),
    written as in form~(\ref{eq:cost-tuple-full-form}),
    is linearly (additively) bounded whenever
    each \( e_i \), \( 0 \leq i \leq m \),
    is linearly (additively) bounded.
    Additionally,
    \( \funcinterpret{\afun} \) is bounded by a functional
    \( f \) if both \( \funcinterpret{\afun}^\cost \)
    and \( \funcinterpret{\afun}^\size \) are bounded by
    \( f \).
\end{Def}

In the next lemma,
we collect the appropriate
induced upper-bounds on innermost runtime complexity
given that we can provide bounds to the \cs{} components
of interpretations.

\begin{Lem}\label{lemma:int-bounds-imply-irc-bounds}
    Suppose \( \rules \) is a TRS compatible with a tuple algebra \( (\typeinterpret{\cdot}, \funcinterpret{}) \),
    then:
    \begin{enumerate}[label = (\roman*)]
        \item if,
        for all \( \afun \in \signature \),
        \( \funcinterpret{\afun}^\size \) is logarithmically
        and \( \funcinterpret{\afun}^\cost \) is additively bounded,
        then \( \irc(n) \in \bigO{\log n} \);

        \item if, for all \( \afun \in \signature \),
        \( \funcinterpret{\afun} \) is additively bounded,
        then \( \irc(n) \in \bigO{n} \);
        and

        \item if, for all defined symbols \( \afun \) and constructors \( \consFont{c} \),
        \( \funcinterpret{\consFont{c}} \) is additively
        and \( \funcinterpret{\afun} \) is polynomially bounded,
        then \( \irc(n) \in \bigO{n^k} \), for some \( k \in \Nat \).
    \end{enumerate}
\end{Lem}

\begin{Exm}\label{ex:int-def-symbols}
Let us illustrate this behavior by interpreting functions
from Example~\ref{ex:toy-trs}.
Interpretation for constructors was given in Example~\ref{ex:int-constructors}.
\[
    \begin{array}{lcllcrlcr}
        \funcinterpret{\add} &=&
        \pair{\costWrapper{\fatlambda x y. y + 1}, \fatlambda x y. x + y}
        &\qquad
        \funcinterpret{\sumList} &=&
        \pair{
            \costWrapper{
                \fatlambda \aListVar.
                2 \aListVar_\leng + \aListVar_\leng \aListVar_\mmax
            },
            \fatlambda \aListVar. \aListVar_\leng \aListVar_\mmax
        }
        \\
        \funcinterpret{\minus} &=&
        \pair{\costWrapper{\fatlambda x y. y + 1}, \fatlambda x y. x}
        &\qquad
        \funcinterpret{\rev} &=&
        \pair{
            \costWrapper{\fatlambda \aListVar. \aListVar_\leng +
            \frac{\aListVar_\leng(\aListVar_\leng + 1)}{2} + 1},
            \fatlambda \aListVar. \aListVar
        }
        \\
        \funcinterpret{\quot} &=&
        \pair{\costWrapper{\fatlambda x y. x + xy + 1}, \fatlambda xy. x}
        \\
        \funcinterpret{\append} &=&
        \multicolumn{4}{l}{
            \pair{
                \costWrapper{
                    \fatlambda \aListVar \bListVar. \aListVar_\leng + 1
                },
                \fatlambda \aListVar \bListVar.
                \pair{
                    \aListVar_\leng + \bListVar_\leng,
                    \max(\aListVar_\mmax, \aListVar_\mmax)
                }
            }
        }
    \end{array}
\]

Checking the compatibility of this interpretation is straightforward.
Notice that in each set of rules defining a function \( \afun \) in
Example~\ref{ex:toy-trs} size components are additively and cost
components are polynomially bounded.
By case (b) of Lemma~\ref{lemma:int-bounds-imply-irc-bounds},
we have that \( \irc[\add] \), \( \irc[\append] \), and \( \irc[\minus] \)
are linear.
Quadratic bounds can be derived to \( \irc[\quot] \),
\( \irc[\sumList] \), and \( \irc[\rev] \).

Recall the semantic meaning given to size components,
see Example~\ref{ex:int-constructors},
one can observe that the cost component of interpretations
do not only bound the innermost runtime complexity of \( \rules_\afun \)
but also provide additional information on the role
each size component plays in the rewriting cost.
For instance: the cost of adding two numbers depends
solely on the size of \( \add \)'s second argument;
the cost of summing every element of a list
has a linear dependency on its length and non-linear
dependency on its length and maximum element.
This is particularly useful in program analysis
since one can detect a possible costly operation
by analyzing the shape of interpretations themselves.
\end{Exm}

\section{Automation}\label{sec:hermes-description}

In this section,
we limn a procedure
for finding \cs{} tuple interpretations.
Our goal is to find interpretations that guarantee polynomial bounds
to the runtime complexity of the rewriting system at hand.
Hence,
we have the following conditions:
(i) the interpretation key chosen is over \( \Nat \),
(ii) the size interpretation of constructors is additively bounded,
and (iii) the interpretation of function symbols is polynomially bounded.

\paragraph*{Parametric Interpretations.}
Recall that previously in the paper
we assigned an intuitive meaning for size components.
In a fully automated setting, where no human guidance is allowed,
all sorts \( \asort \) start with \( \typecount{\asort} = 1 \)
and go up to a predefined bound \( K \).
This maximum bound \( K \) is needed to limit the search space
and guarantee that the procedure terminates.

Roughly, the procedure works as follows.
The interpretation of data constructors is set to be additive.
So if
\( \consFont{c} : \asort_1 \arrtype \dots \arrtype \asort_m \arrtype \bsort \)
is a data constructor,
its size interpretation is
\(
    \fatlambda x_1 \dots x_m.\,
    a + \sum_{i = 1}^m\sum_{j = 1}^{\typecount{\asort_i}}
    x_{ij}
\),
where \( a \) is a parameter to be determined by the search procedure.
We say such an \textit{interpretation shape}
is parametrized by the coefficient \( a \).
The next step is to choose (parametric) interpretations for defined symbols
\( \afun \in \signature \).
In contrast with constructors where the cost components are zero-valued functions
and size components are additive,
we can choose any function that is polynomially bounded for
cost and size components of a defined symbol \( \afun \in \signature \).

However, the class of functions from which we can choose interpretations
of defined symbols is too big.
So we restrict our search space to a limited class of polynomially
bounded functions: max-polynomials, i.e.,
functions that combine polynomial terms and the \( \max \) function.
For instance,
the interpretations of \( \cons \) in Example~\ref{ex:int-constructors}
and \( \append \) in Example~\ref{ex:int-def-symbols} are max-polynomials.
We then choose generic max-polynomials for the cost and size components
which are parametrized by their coefficients.
Recall that we wish for finding interpretations that satisfy
the compatibility condition, i.e.,
\( \interpret{\ell}_\alpha \cartGt \interpret{r}_\alpha \),
for any \( \alpha \).
Therefore,
if we pick max-polynomials parametrized by their coefficients,
those give rise to a set of constraints that must be solved in order
to determine valid interpretations.

\begin{Exm}\label{ex:find-int-double}
    Let us illustrate the ideas above with a simple system defining the
    function \( \double \) over natural numbers.
    So we consider the system with rules \( \double \app \zero \arrz \zero \) and
    \( \double \app (\suc \app \avar) \arrz
    \suc \app (\suc \app (\double \app \avar)) \).
    Let us choose the following parametric interpretation
    \begin{align*}
        \funcinterpret{\zero} &=
        \pair{0, a_0}
        &
        \funcinterpret{\suc} &=
        \pair{\costWrapper{\fatlambda x. 0}, \fatlambda x. x + b_0}
        &
        \funcinterpret{\double} &=
        \pair{\costWrapper{\fatlambda x. c_1 x + c_0}, \fatlambda x. d_1 x + d_0},
    \end{align*}
    which satisfy conditions (i)-(iii) above.
    The interpretation above is \textit{parametric} in the sense that
    the coefficients
    \( a_0, b_0, c_0, c_1, d_0, d_1 \)
    are to be determined.
    The compatibility condition for the first rule gives:
    \[
        \interpret{\double \app \zero} \cartGt \interpret{\zero}
        \implies
        \pair{\costWrapper{c_1 a_0 + c_0}, d_1 a_0 + d_0}
        \cartGt
        \pair{0, a_0},
    \]
    which in consequence requires the validity of
    \( C_0 = (c_1 a_0 + c_0 > 0) \wedge (d_1 a_0 + d_0 \geq a_0) \).
    The compatibility condition for the second rule, on the other hand,
    gives us the following:
    \[
        \interpret{\double \app (\suc \app \avar)}
        \cartGt
        \interpret{\suc \app (\suc \app (\double \app \avar))}
        \implies
        \pair{\costWrapper{c_1 x + c_1 b_0 + c_0}, d_1 x + d_1 b_0 + d_0}
        \cartGt
        \pair{\costWrapper{c_1 x + c_0}, d_1 x + d_0 + 2 b_0},
    \]
    which in consequence requires the validity of the formula
    \[
        C_1 = (c_1 x + c_1 b_0 + c_0 > c_1 x + c_0)
        \wedge
        (d_1 x + d_1 b_0 + d_0 \geq d_1 x + d_0 + 2 b_0).
    \]
    Hence,
    we seek to find witnesses for the constraints \( C_0, C_1 \)
    over \( \Nat \).
    For which we can use an SMT solver.
\end{Exm}

The example above is very simple in nature
but uses the main ideas of our procedure.
Essentially,
we choose parametric interpretations for function symbols in \( \signature \)
and solve the constraints that arise from the compatibility condition.
As we have seen in Example~\ref{ex:int-def-symbols},
\cs{} interpretations may become complicated,
so more interpretation shapes are needed in the search procedure.
We describe such a procedure below.
It is modular in the sense that it is parametrized
by a \textit{selector strategy} \( \mathcal{S} \) and constraint solver.
A selector strategy is an algorithm to choose a parametric interpretation for each
function symbol in \( \signature \).
For instance,
in the example above we have chosen
\textit{linear parametric shapes} for all function symbol.

\bigskip
\hrule
\smallskip
\noindent\textbf{Main Procedure}
\smallskip
\hrule
\smallskip
\noindent\textbf{Parameter:} A selector algorithm \( \mathcal{S} \)
and a constraint solver over non-linear integer arithmetic.\\
\noindent\textbf{Data Input:}
A TRS \( \rules \) over a syntax signature \( (\sortset, \signature, \ar) \).\\
\textbf{Output:}
\( \mathtt{YES} \),
if a \cs{} tuple interpretation satisfying compatibility can be found and
\( \mathtt{MAYBE} \),
if all steps below were executed and no interpretation
could be found\footnote{Notice that in our setting we cannot possibly return \( \mathtt{NO} \).}.
\begin{enumerate}
    \item Split \( \signature \) into two disjoint sets of constructors and defined symbols,
    i.e., \( \signature = \ctrS \uplus \dfdS \).

    \item For each constructor \( \consFont{c} \hasType \asort_1 \arrtype \dots \arrtype \asort_m \arrtype \bsort \),
    choose its cost interpretation as the zero-valued cost function;
    size interpretations are additive.

    \item Split \( \dfdS \) into sets \( \dfdS_1, \dots, \dfdS_n \)
    such that for each \( \afun \in \dfdS_i \),
    with \( 1 \leq i \leq n \),
    all function symbols occurring in the rules defining \( \afun \) are either constructors or
    in \( \dfdS_1 \cup \dots \cup \dfdS_i \).

    \item For each \( 1 \leq i \leq n \),
        choose an \emph{interpretation shape} for the symbols in \( \dfdS_i \) based
        on the selector strategy \( \mathcal{S} \) (to be defined below).
    \begin{itemize}
        \item Mark the chosen interpretation shape on \( \mathcal{S} \),
        so we don't choose the same again in case this step fails.
        \item If no choice can be made by \( \mathcal{S} \),
        stop and return \texttt{MAYBE}\@.
    \end{itemize}

    \item
    If \( \afun \app \ell_1 \app \dots \app \ell_k \arrz r \)
    is a rule of type \( \asort \) with
    \( \afun \in \mathcal{D}_1 \cup \dots \cup \mathcal{D}_i \).
    \emph{Simplify}
    \( \interpret{\afun \app \ell_1 \app \dots \app \ell_k}
    \cartGt \interpret{r} \) so that
    the result is a set of inequality constraints \( C \)
    that does not depend on any interpreted variable
    (we shall define this simplification step below).

    \begin{itemize}
        \item If this simplification step fails,
        then we return to step 4 to choose another interpretation shape.
    \end{itemize}

    \item \emph{Check} if \( C \) holds.
    \begin{itemize}
        \item If all constraints in \( C \) hold and \( i < n \),
        it means that we could orient all rules headed by function symbols
        in \( \mathcal{D}_i \), so we go to step 4 with \( i := i + 1 \).
        \item If all constraints in \( C \) hold and \( i = n \),
        then we could orient all rules \( \rules \),
        stop return \texttt{YES}.
        \item Otherwise,
        increase \( \typecount{\asort} \) by one,
        update the additive size interpretation for the constructors,
        and return to step 4 choosing another interpretation shape.
    \end{itemize}
\end{enumerate}
\smallskip
\hrule
\bigskip

Two key aspects of the procedure above remain to be defined.
The strategy \( \mathcal{S} \) for selecting interpretation shapes
and the constraint solver, Step 6.

\paragraph{Strategy-based Search for Tuple Interpretations.}

Intuitively, a selector strategy \( \mathcal{S} \)
is an algorithm for choosing parametric interpretations
for defined symbols in \( \dfdS_i \).
For instance,
we could randomly pick an interpretation shape from a list
(the \textbf{blind} strategy);
we could incrementally select interpretations from a list
of possible attempts
(the \textbf{progressive} strategy);
or we could select interpretations based on their syntax patterns
(the \textbf{pattern} strategy).

The definition below lists some interpretation shapes we consider.
They are based on
the classes studied in~\cite{steinbach:05,contejan:tomas:urbain:05}
Parametric interpretations are built by considering the type of defined symbols.

\begin{Def}[Interpretation Shapes]\label{def:int-shapes}
    Let \( \atype = \asort_1 \arrtype \dots \arrtype \asort_m \arrtype \bsort \)
    and each \( f_{ij} \) appearing in the shapes
    below be an additively bounded weakly monotonic
    function over \( \sizeInt{\atype} \).
    We write \( f(\vec{x}) \) for the application of \( f \) to each argument
    \( x_1, \dots, x_m \).
    \begin{itemize}
        \item
        The \emph{additive class} contains additively bounded \cs{} functionals
        of the following form:
        \[
            \fatlambda x_1 \dots x_m.\;
                {\sum_{i = 1}^m\sum_{j = 1}^{\typecount{\asort_i}} x_{ij}}
                + b_0 + f(\vec{x})
        \]

        \item The \emph{linear class} contains \cs{} functionals written as:
        \[
            \fatlambda x_1 \dots x_m.\;
                {\sum\limits_{i = 1}^m
                \sum\limits_{j = 1}^{\typecount{\asort_i}}}
                {a_{ij} x_{ij}} {f_{ij}(\vec{x}) }
        \]

        \item The \emph{simple class} contains \cs{} functionals written as:
        \[
            \fatlambda x_1 \dots x_m.\;
                {\sum\limits_{i = 1}^m
                \sum\limits_{j = 1}^{\typecount{\asort_i}}}
                {a_{ij} x_{ij}^{k_{ij}}} {f_{ij}(\vec{x}) },
                \text{ such that each } k_{ij} \in \{ 0, 1 \}
        \]

        \item
        Finally,
        the \emph{quadratic} class contains \cs{} functionals
        where we allow general products of variables
        with degree at maximum \( 2 \):
        \[
            \fatlambda x_1 \dots x_m.\;
                {\sum\limits_{i = 1}^m
                \sum\limits_{j = 1}^{\typecount{\asort_i}}}
                {a_{ij} x_{ij}^{k_{ij}}} {f_{ij}(\vec{x}) },
                \text{ such that each } k_{ij} \in \{ 0, 1, 2 \}
        \]

        \item The \emph{simple quadratic} class contains \cs{}
        functionals built as a sum of a simple functional
        plus a quadratic component:
        \[
            \fatlambda x_1 \dots x_m.\;
                {\sum\limits_{i = 1}^m
                \sum\limits_{j = 1}^{\typecount{\asort_i}}}
                {a_{ij} x_{ij}^{k_{ij}}} {f_{ij}(\vec{x}) }
                +
                {\sum\limits_{i = 1}^m
                \sum\limits_{j = 1}^{\typecount{\asort_i}}}
                {a_{ij} x_{ij}^{l_{ij}}} {f_{ij}(\vec{x}) },
        \]
        with \( k_{ij} \in \{ 0,1\} \) and \( l_{ij} \in \{ 0,1,2 \} \).
    \end{itemize}
\end{Def}

Hence,
the blind strategy randomly selects one of the shapes above.
The incremental strategy chooses interpretations in order,
from additive ones to quadratic ones.
The pattern strategy is slightly more difficult to realize
since we need heuristic analysis on the shape of rules.
For instance,
every rule of the form
\( \afun \app x_1 \dots x_m \arrz x_i \)
have constant cost functions
\( \costWrapper{\fatlambda x_1 \dots x_m. 1} \)
and additive size components.
Rules that duplicate variables,
as in the pattern
\( C[x] \arrz D[x,x] \),
induce at least quadratic bound on cost.
Notice that this is the case for all quadratic
complexities in this paper.
The concrete implementation of a selector algorithm determines the efficiency
of the main procedure for finding interpretations.

In order to simplify constraints \( \interpret{\ell} \cartGt \interpret{r} \)
we have to simplify inequalities between polynomials (max-polynomials).
To simplify polynomial (max-polynomial) shapes,
 we need to compare polynomials
\( P_\ell^\cost \costGt R_r^\cost \) and
\( P_{\ell_1}^\size \sizeGe P_{r_1}^\size
\wedge
\dots
\wedge
P_{\ell_{\typecount{\btype}}}^\size
\sizeGe
P_{r_{\typecount{\bsort}}}^\size
 \).
These conditions are then reduced to formulas in \textsf{QF_NIA} (Quantifier-Free Non-Linear Integer Arithmetic)
and sent to an SMT solver, see~\cite{giesl:et-al:17}.
Max-polynomials are simplified using the rules
\( \max(x,y) + z \leadsto \max(x + z, y + z) \) and \( \max(x,y) z \leadsto \max(x z, y z) \).
The result has the form \( \max_l P_l \) where each \( P_l \) is a polynomial without max
occurrences~\cite{codish:gonopolskiy:ben-amram:amir:fuhs:giesl:11}.

\section{Conclusion}\label{sec:conc}

In this paper
we showed that \cs{} tuple pairs
can be adapted to handle innermost rewriting.
The type-aware algebraic interpretation style
provided the machinery necessary to deal with innermost
termination and
a mechanism to establish upper bounds to the innermost runtime complexity of compatible TRSs.
We presented sufficient conditions for feasible (polynomial)
bounds on \( \irc \) of compatible systems,
which are in line with related works in the literature.
This line of investigation is far from over.
Since searching for interpretations can be cumbersome,
our immediate future work is to
develop new strategies and interpretation shapes.
For instance,
we seek to expand the class of interpretations beyond max-polynomials
such as logarithmic functionals.
This has the potential to drastically improve the efficiency
of our tooling.

\paragraph{Acknowledgments.}
We wish to thank Cynthia Kop ---
for the valuable discussions and guidance
during the production of this paper;
we thank
Niels van der Weide, Marcos Bueno, and Edna Gomes ---
for carefully proofreading the various
manuscript versions of the paper;
and we thank the anonymous referees ---
for their valuable comments that helped us improve the paper.

\bibliographystyle{eptcs}
\bibliography{references}

\begin{thebibliography}{10}
\providecommand{\bibitemdeclare}[2]{}
\providecommand{\surnamestart}{}
\providecommand{\surnameend}{}
\providecommand{\urlprefix}{Available at }
\providecommand{\url}[1]{\texttt{#1}}
\providecommand{\href}[2]{\texttt{#2}}
\providecommand{\urlalt}[2]{\href{#1}{#2}}
\providecommand{\doi}[1]{doi:\urlalt{https://doi.org/#1}{#1}}
\providecommand{\eprint}[1]{arXiv:\urlalt{https://arxiv.org/abs/#1}{#1}}
\providecommand{\bibinfo}[2]{#2}

\bibitemdeclare{inproceedings}{ava:mos:08}
\bibitem{ava:mos:08}
\bibinfo{author}{M.~\surnamestart Avanzini\surnameend} \&
  \bibinfo{author}{G.~\surnamestart Moser\surnameend} (\bibinfo{year}{2008}):
  \emph{\bibinfo{title}{Complexity Analysis by Rewriting}}.
\newblock In: {\slshape \bibinfo{booktitle}{Proc.\ FLOPS}}, pp.
  \bibinfo{pages}{130--146}, \doi{10.1007/978-3-540-78969-7_11}.

\bibitemdeclare{article}{baillot:lago:16}
\bibitem{baillot:lago:16}
\bibinfo{author}{P.~\surnamestart Baillot\surnameend} \&
  \bibinfo{author}{U.~\surnamestart Dal~Lago\surnameend}
  (\bibinfo{year}{2016}): \emph{\bibinfo{title}{Higher-order interpretations
  and program complexity}}.
\newblock {\slshape \bibinfo{journal}{IC}}, pp. \bibinfo{pages}{56--81},
  \doi{10.1016/j.ic.2015.12.008}.

\bibitemdeclare{article}{cherifa:lescanne:87}
\bibitem{cherifa:lescanne:87}
\bibinfo{author}{A.~\surnamestart {Ben Cherifa}\surnameend} \&
  \bibinfo{author}{P.~\surnamestart Lescanne\surnameend}
  (\bibinfo{year}{1987}): \emph{\bibinfo{title}{Termination of rewriting
  systems by polynomial interpretations and its implementation}}.
\newblock {\slshape \bibinfo{journal}{Science of Computer Programming}}
  \bibinfo{volume}{9}(\bibinfo{number}{2}), pp. \bibinfo{pages}{137--159},
  \doi{10.1016/0167-6423(87)90030-X}.

\bibitemdeclare{article}{bonfante:cichon:marion:touzet:01}
\bibitem{bonfante:cichon:marion:touzet:01}
\bibinfo{author}{G.~\surnamestart Bonfante\surnameend},
  \bibinfo{author}{A.~\surnamestart Cichon\surnameend}, \bibinfo{author}{J.-Y.
  \surnamestart Marion\surnameend} \& \bibinfo{author}{H.~\surnamestart
  Touzet\surnameend} (\bibinfo{year}{2001}): \emph{\bibinfo{title}{Algorithms
  with polynomial interpretation termination proof}}.
\newblock {\slshape \bibinfo{journal}{Journal of Functional Programming}}
  \bibinfo{volume}{11}(\bibinfo{number}{1}), p. \bibinfo{pages}{33–53},
  \doi{10.1017/S0956796800003877}.

\bibitemdeclare{inproceedings}{bonfante:et-al:01}
\bibitem{bonfante:et-al:01}
\bibinfo{author}{G.~\surnamestart Bonfante\surnameend},
  \bibinfo{author}{J.~\surnamestart Marion\surnameend} \&
  \bibinfo{author}{J.~\surnamestart Moyen\surnameend} (\bibinfo{year}{2001}):
  \emph{\bibinfo{title}{On Lexicographic Termination Ordering with Space Bound
  Certifications}}.
\newblock In: {\slshape \bibinfo{booktitle}{Proc.\ PSI}}, pp.
  \bibinfo{pages}{482--493}, \doi{10.1007/3-540-45575-2_46}.

\bibitemdeclare{inproceedings}{cichon:lescanne:92}
\bibitem{cichon:lescanne:92}
\bibinfo{author}{A.~\surnamestart Cichon\surnameend} \&
  \bibinfo{author}{P.~\surnamestart Lescanne\surnameend}
  (\bibinfo{year}{1992}): \emph{\bibinfo{title}{Polynomial interpretations and
  the complexity of algorithms}}.
\newblock In: {\slshape \bibinfo{booktitle}{CADE}}, pp.
  \bibinfo{pages}{139--147}, \doi{10.1007/3-540-55602-8_161}.

\bibitemdeclare{article}{codish:gonopolskiy:ben-amram:amir:fuhs:giesl:11}
\bibitem{codish:gonopolskiy:ben-amram:amir:fuhs:giesl:11}
\bibinfo{author}{M.~\surnamestart Codish\surnameend},
  \bibinfo{author}{I.~\surnamestart Gonopolskiy\surnameend},
  \bibinfo{author}{A.~M. \surnamestart Ben-Amram\surnameend},
  \bibinfo{author}{C.~\surnamestart Fuhs\surnameend} \&
  \bibinfo{author}{J.~\surnamestart Giesl\surnameend} (\bibinfo{year}{2011}):
  \emph{\bibinfo{title}{SAT-based termination analysis using monotonicity
  constraints over the integers}}.
\newblock {\slshape \bibinfo{journal}{Theory and Practice of Logic
  Programming}} \bibinfo{volume}{11}(\bibinfo{number}{4-5}), p.
  \bibinfo{pages}{503–520}, \doi{10.1017/S1471068411000147}.

\bibitemdeclare{article}{contejan:marche:tomas:urbain:05}
\bibitem{contejan:marche:tomas:urbain:05}
\bibinfo{author}{E.~\surnamestart Contejan\surnameend},
  \bibinfo{author}{C.~\surnamestart Marché\surnameend}, \bibinfo{author}{A.~P.
  \surnamestart Tomás\surnameend} \& \bibinfo{author}{X.~\surnamestart
  Urbain\surnameend} (\bibinfo{year}{2005}): \emph{\bibinfo{title}{Mechanically
  Proving Termination Using Polynomial Interpretations}}.
\newblock {\slshape \bibinfo{journal}{JAR}}
  \bibinfo{volume}{34}(\bibinfo{number}{34}), \doi{10.1007/s10817-005-9022-x}.

\bibitemdeclare{article}{contejan:tomas:urbain:05}
\bibitem{contejan:tomas:urbain:05}
\bibinfo{author}{Evelyne \surnamestart Contejean\surnameend},
  \bibinfo{author}{Claude \surnamestart March{\'e}\surnameend},
  \bibinfo{author}{Ana~Paula \surnamestart Tom{\'a}s\surnameend} \&
  \bibinfo{author}{Xavier \surnamestart Urbain\surnameend}
  (\bibinfo{year}{2005}): \emph{\bibinfo{title}{Mechanically Proving
  Termination Using Polynomial Interpretations}}.
\newblock {\slshape \bibinfo{journal}{JAR}}
  \bibinfo{volume}{34}(\bibinfo{number}{4}), pp. \bibinfo{pages}{325--363},
  \doi{10.1007/s10817-005-9022-x}.

\bibitemdeclare{article}{giesl:et-al:17}
\bibitem{giesl:et-al:17}
\bibinfo{author}{J.~\surnamestart Giesl\surnameend},
  \bibinfo{author}{C.~\surnamestart Aschermann\surnameend},
  \bibinfo{author}{M.~\surnamestart Brockschmidt\surnameend},
  \bibinfo{author}{F.~\surnamestart Emmes\surnameend},
  \bibinfo{author}{F.~\surnamestart Frohn\surnameend},
  \bibinfo{author}{C.~\surnamestart Fuhs\surnameend},
  \bibinfo{author}{J.~\surnamestart Hensel\surnameend},
  \bibinfo{author}{C.~\surnamestart Otto\surnameend},
  \bibinfo{author}{M.~\surnamestart Plucker\surnameend},
  \bibinfo{author}{P.~\surnamestart Schneider-Kamp\surnameend},
  \bibinfo{author}{T.~\surnamestart Stroder\surnameend},
  \bibinfo{author}{S.~\surnamestart Swiderski\surnameend} \&
  \bibinfo{author}{R.~\surnamestart Thiemann\surnameend}
  (\bibinfo{year}{2017}): \emph{\bibinfo{title}{Analyzing Program Termination
  and Complexity Automatically with \textsf{AProVE}}}.
\newblock {\slshape \bibinfo{journal}{JAR}} \bibinfo{volume}{58}, pp.
  \bibinfo{pages}{3--31}, \doi{10.1007/s10817-016-9388-y}.

\bibitemdeclare{inproceedings}{nao:moser:08}
\bibitem{nao:moser:08}
\bibinfo{author}{N.~\surnamestart Hirokawa\surnameend} \&
  \bibinfo{author}{G.~\surnamestart Moser\surnameend} (\bibinfo{year}{2008}):
  \emph{\bibinfo{title}{Automated Complexity Analysis Based on the Dependency
  Pair Method}}.
\newblock In: {\slshape \bibinfo{booktitle}{Proc.\ IJCAR}}, pp.
  \bibinfo{pages}{364--379}, \doi{10.1007/978-3-540-71070-7_32}.

\bibitemdeclare{article}{hof:92}
\bibitem{hof:92}
\bibinfo{author}{D.~\surnamestart Hofbauer\surnameend} (\bibinfo{year}{1992}):
  \emph{\bibinfo{title}{Termination proofs by multiset path orderings imply
  primitive recursive derivation lengths}}.
\newblock {\slshape \bibinfo{journal}{Proc.\ TCS}},
  \doi{10.1007/3-540-53162-9_50}.

\bibitemdeclare{inproceedings}{hof:01}
\bibitem{hof:01}
\bibinfo{author}{D.~\surnamestart Hofbauer\surnameend} (\bibinfo{year}{2001}):
  \emph{\bibinfo{title}{Termination Proofs by Context-Dependent
  Interpretations}}.
\newblock In: {\slshape \bibinfo{booktitle}{Proc.\ RTA}}, pp.
  \bibinfo{pages}{108--121}, \doi{10.1007/3-540-45127-7_10}.

\bibitemdeclare{inproceedings}{hof:lau:89}
\bibitem{hof:lau:89}
\bibinfo{author}{D.~\surnamestart Hofbauer\surnameend} \&
  \bibinfo{author}{C.~\surnamestart Lautemann\surnameend}
  (\bibinfo{year}{1989}): \emph{\bibinfo{title}{Termination proofs and the
  length of derivations}}.
\newblock In: {\slshape \bibinfo{booktitle}{Proc.\ RTA}}, pp.
  \bibinfo{pages}{167--177}, \doi{10.1007/3-540-51081-8_107}.

\bibitemdeclare{book}{huet:oppen:80}
\bibitem{huet:oppen:80}
\bibinfo{author}{G.~\surnamestart Huet\surnameend} \& \bibinfo{author}{D.C
  \surnamestart Oppen\surnameend} (\bibinfo{year}{1980}):
  \emph{\bibinfo{title}{Equations and rewrite rules: a survey}}.
\newblock \bibinfo{series}{Formal Language Theory: Perspectives and Open
  Problems}, \bibinfo{publisher}{Loria}.
\newblock \urlprefix\url{http://rewriting.loria.fr/documents/CS-TR-80-785.pdf}.

\bibitemdeclare{inproceedings}{kop:vale:21}
\bibitem{kop:vale:21}
\bibinfo{author}{C.~\surnamestart Kop\surnameend} \&
  \bibinfo{author}{D.~\surnamestart Vale\surnameend} (\bibinfo{year}{2021}):
  \emph{\bibinfo{title}{{Tuple Interpretations for Higher-Order Complexity}}}.
\newblock In: {\slshape \bibinfo{booktitle}{FSCD}}, pp.
  \bibinfo{pages}{31:1--31:22}, \doi{10.4230/LIPIcs.FSCD.2021.31}.

\bibitemdeclare{article}{lautemann:88}
\bibitem{lautemann:88}
\bibinfo{author}{C.~\surnamestart Lautemann\surnameend} (\bibinfo{year}{1988}):
  \emph{\bibinfo{title}{A note on polynomial interpretation}}.
\newblock {\slshape \bibinfo{journal}{Bulletin EATCS}} \bibinfo{volume}{volume
  36}, pp. \bibinfo{pages}{129--131}.

\bibitemdeclare{inproceedings}{mitterwallner:aart:22}
\bibitem{mitterwallner:aart:22}
\bibinfo{author}{F.~\surnamestart Mitterwallner\surnameend} \&
  \bibinfo{author}{A.~\surnamestart Middeldorp\surnameend}
  (\bibinfo{year}{2022}): \emph{\bibinfo{title}{{Polynomial Termination Over
  \(\mathbb{N}\) Is Undecidable}}}.
\newblock In: {\slshape \bibinfo{booktitle}{Proc.\ FSCD}}, pp.
  \bibinfo{pages}{27:1--27:17}, \doi{10.4230/LIPIcs.FSCD.2022.27}.

\bibitemdeclare{inproceedings}{moser:17}
\bibitem{moser:17}
\bibinfo{author}{G.~\surnamestart Moser\surnameend} (\bibinfo{year}{2017}):
  \emph{\bibinfo{title}{{Uniform Resource Analysis by Rewriting: Strengths and
  Weaknesses (Invited Talk)}}}.
\newblock In: {\slshape \bibinfo{booktitle}{Proc.\ FSCD}}, pp.
  \bibinfo{pages}{2:1--2:10}, \doi{10.4230/LIPIcs.FSCD.2017.2}.

\bibitemdeclare{inproceedings}{moser:schnabl:wadmann:08}
\bibitem{moser:schnabl:wadmann:08}
\bibinfo{author}{G.~\surnamestart Moser\surnameend},
  \bibinfo{author}{A.~\surnamestart Schnabl\surnameend} \&
  \bibinfo{author}{J.~\surnamestart Waldmann\surnameend}
  (\bibinfo{year}{2008}): \emph{\bibinfo{title}{{Complexity Analysis of Term
  Rewriting Based on Matrix and Context Dependent Interpretations}}}.
\newblock In: {\slshape \bibinfo{booktitle}{Proc.\ IARCS}}, pp.
  \bibinfo{pages}{304--315}, \doi{10.4230/LIPIcs.FSTTCS.2008.1762}.

\bibitemdeclare{inproceedings}{noschinski:emmes:jurgen:11}
\bibitem{noschinski:emmes:jurgen:11}
\bibinfo{author}{L.~\surnamestart Noschinski\surnameend},
  \bibinfo{author}{F.~\surnamestart Emmes\surnameend} \&
  \bibinfo{author}{J.~\surnamestart Giesl\surnameend} (\bibinfo{year}{2011}):
  \emph{\bibinfo{title}{A Dependency Pair Framework for Innermost Complexity
  Analysis of Term Rewrite Systems}}.
\newblock In: {\slshape \bibinfo{booktitle}{CADE-23}}, pp.
  \bibinfo{pages}{422--438}, \doi{10.1007/978-3-642-22438-6_32}.

\bibitemdeclare{inproceedings}{steinbach:05}
\bibitem{steinbach:05}
\bibinfo{author}{J.~\surnamestart Steinbach\surnameend} (\bibinfo{year}{1992}):
  \emph{\bibinfo{title}{Proving polynomials positive}}.
\newblock In: {\slshape \bibinfo{booktitle}{In Proc.\ FSTTCS 92}},
  \bibinfo{address}{Berlin, Heidelberg}, pp. \bibinfo{pages}{191--202},
  \doi{10.1007/3-540-56287-7_105}.

\bibitemdeclare{article}{toyama:87}
\bibitem{toyama:87}
\bibinfo{author}{Yoshihito \surnamestart T.\surnameend} (\bibinfo{year}{1987}):
  \emph{\bibinfo{title}{Counterexamples to termination for the direct sum of
  term rewriting systems}}.
\newblock {\slshape \bibinfo{journal}{Information Processing Letters}}
  \bibinfo{volume}{25}(\bibinfo{number}{3}), pp. \bibinfo{pages}{141--143},
  \doi{10.1016/0020-0190(87)90122-0}.

\bibitemdeclare{article}{weiermann:95}
\bibitem{weiermann:95}
\bibinfo{author}{A.~\surnamestart Weiermann\surnameend} (\bibinfo{year}{1995}):
  \emph{\bibinfo{title}{Termination proofs for term rewriting systems by
  lexicographic path orderings imply multiply recursive derivation lengths}}.
\newblock {\slshape \bibinfo{journal}{TCS}},
  \doi{10.1016/0304-3975(94)00135-6}.

\bibitemdeclare{article}{yamada:22}
\bibitem{yamada:22}
\bibinfo{author}{A.~\surnamestart Yamada\surnameend} (\bibinfo{year}{2022}):
  \emph{\bibinfo{title}{Tuple Interpretations for Termination of Term
  Rewriting}}.
\newblock {\slshape \bibinfo{journal}{J Autom Reasoning}},
  \doi{10.1007s10817-022-09640-4}.

\end{thebibliography}

\end{document}